\documentclass[]{JFM-FLM_Au}

\usepackage[table,dvipsnames]{xcolor}
\usepackage{placeins}

\lefttitle{Covarrubias et al.}
\righttitle{Journal of Fluid Mechanics}

\title{Solute pulse compression and spreading in power-law
fluid flow through a semipermeable tube}

\author{Geraldy Covarrubias\aff{1}, José Arcos\aff{1},  Federico Méndez\aff{2}, \and Oscar Bautista\aff{1}}

\affiliation{\aff{1}ESIME Azcapotzalco, Instituto Politécnico Nacional, México City, 02250, Mexico
\aff{2} Departamento de Termofluidos, Facultad de Ingeniería, Universidad Nacional Autónoma de México, México City, 04510, Mexico}

\corresau{O. Bautista, obautista@ipn.mx}

\begin{document}
\maketitle

\begin{abstract}
We investigate Taylor–Aris dispersion and filtration-induced compression of a solute pulse transported by a power-law fluid through a slender, semipermeable, circular tube. The wall permits solvent filtration, modelled by Starling’s law, while completely retaining the solute. Lubrication theory and multiple-scale homogenisation yield a one-dimensional macrotransport equation with spatially varying effective advection and dispersion coefficients. An evolution equation for the axial variance identifies the competing contributions of mean-flow deceleration, dispersive spreading, and wall-induced concentration redistribution. Filtration causes the mean axial velocity to decrease downstream, allowing compressive advection to overcome dispersive spreading. A local critical-width criterion distinguishes instantaneous pulse contraction from spreading. We evaluate the contraction threshold using the integral variance balance and compare the predictions with axisymmetric convection–diffusion simulations. As the relative contributions of compression and dispersion evolve, an initially contracting pulse may reach a minimum width and subsequently spread. Power-law rheology modifies both the hydraulic field and the Taylor dispersion contribution. Under the reference conditions investigated, shear thinning produces greater cumulative pulse narrowing than the Newtonian case: at dimensionless time $T=100$, the pulse width decreases from its initial value by approximately $31\%$ for flow behaviour index $n=0.4$, compared with $26\%$ for $n=1$. Substantial pulse narrowing can coexist with an outlet solvent flow rate that is only a small fraction of the inlet value.
\end{abstract}


\section{Introduction}
Solute transport in confined flows is governed by the interplay between axial advection, transverse molecular diffusion and spatial variations in the velocity field. These mechanisms determine how concentration distributions evolve in microfluidic, physiological, and environmental flows. In chemical separation systems and microfluidic devices, dispersion analysis helps predict solute-band broadening and assess how channel geometry affects transport \citep{ChangSantiago2023}. In heterogeneous porous media, it is used to characterise solute spreading and its dependence on fluid rheology \citep{an2022upscaling,omrani2024anomalies,alqenae2025enhanced}. Physiological transport and drug delivery also motivate understanding how flow redistributes dissolved substances and alters localised concentration profiles \citep{grotberg2004biofluid}. These applications motivate a quantitative description of the competition between dispersive spreading and flow-induced pulse compression. In pressure-driven flow, velocity shear interacts with transverse molecular diffusion to enhance axial spreading. Classical analyses by \citet{taylor1953dispersion} and \citet{aris1956dispersion} describe this mechanism using an effective dispersion coefficient and concentration moments. However, spatially varying advection can counteract dispersive spreading and produce transient pulse contraction.

Conversely, antidispersion can narrow and concentrate solute distributions through selective solvent extraction across semipermeable walls, offering a potential mechanism to control solute concentration in microfluidic and biomedical systems \citep{gan2025antidispersion}. In this context, they investigated \emph{antidispersion} of a retained solute in a Newtonian fluid flowing between semipermeable parallel plates. Solvent filtration reduces the mean downstream axial velocity, allowing compressive advection to overcome dispersive spreading. Here we extend this mechanism to a circular tube carrying a power-law fluid, whose rheology changes both the velocity profile and the axial hydrodynamic response. We therefore examine how power-law rheology modifies the balance between filtration-induced compression and dispersive spreading.

Classical dispersion theory has also been extended to non-Newtonian fluids, for which the constitutive response modifies the velocity profile and its contribution to axial spreading \citep{sharp1993shear}. For power-law fluids, shear-thinning behaviour ($n<1$, where $n$ is the flow behaviour index) flattens the velocity profile and reduces the profile-dependent Taylor contribution when the mean velocity and molecular diffusivity are held fixed. Nevertheless, the effect of shear thinning on macroscopic spreading depends on the flow geometry and operating protocol. Simulations and experiments in heterogeneous porous media report non-monotonic dispersivity and, in some cases, enhanced spreading relative to Newtonian controls \citep{an2022upscaling,omrani2024anomalies,alqenae2025enhanced}. Although these geometries differ from the present tube configuration, they highlight the need to distinguish changes in Taylor dispersion arising from the velocity-profile shape at fixed mean velocity from those associated with the coupled hydraulic response.

Wall interactions further modify the results: for example, \citet{rana2016taylor} studied solute dispersion in pulsatile Casson fluid flow through a tube with wall absorption. More recently, \citet{singh2023skewness} analysed pulsatile Carreau--Yasuda flow with wall absorption using concentration moments up to fourth order, including skewness, kurtosis, and departures from Gaussian concentration profiles. These studies connect rheology and wall interactions to both dispersion coefficients and finite-time pulse shape. Porous-wall dispersion models also include radial advection, diffusion, and solute uptake \citep{griffiths2013control}, as well as power-law rheology \citep{dejam2018dispersion}. In the latter study, concentration and diffusive flux are continuous between the conduit and an adjacent porous medium, allowing solute exchange. These conditions differ from the boundary condition considered here, in which solvent crosses the wall but the solute is completely retained.

Advection also varies spatially in channels with varying cross-section. \citet{ChangSantiago2023} developed a Taylor--Aris dispersion model for axisymmetric channels with a slowly varying radius, derived evolution equations for the solute centroid and axial variance, and identified conditions for transient pulse contraction. The relevance of tracking solute moments beyond the long-time dispersion coefficient was also demonstrated by \citet{samuel2025electroosmotic}, who analysed coupled electroosmotic and pressure-driven flows across all time regimes. For non-Newtonian transport, \citet{teodoro2025axisymmetric} studied Taylor--Aris dispersion of a simplified Phan--Thien--Tanner fluid in axisymmetric channels with small axial wall slopes and identified conditions for zero or negative transient variance growth. More recently, \citet{samuel2026geometry} developed an area-averaged transport framework for long channels with arbitrary, gradually varying cross-sectional size and shape, along with equations for the solute centroid and axial variance, and comparisons with Brownian dynamics simulations. These studies motivate a moment-based assessment of pulse evolution. In the present constant-radius tube, axial flow non-uniformity instead results from solvent filtration through a solute-retaining wall.

The one-dimensional Taylor--Aris description is an asymptotic reduction that requires sufficient transverse mixing. As \citet{young1991shear} discuss, the evolution before this regime is reached can retain transient features absent from the classical reduced equation. More recently, \citet{taghizadeh2020preasymptotic} developed an averaged description that accounts explicitly for relaxation from non-uniform initial concentration fields. These considerations highlight the importance of distinguishing initial cross-sectional relaxation from compression induced by spatially varying advection. They also motivate comparison of the homogenised model with cross-sectionally averaged solutions of the axisymmetric transport equation.

A further constitutive issue concerns molecular diffusion itself. \citet{boyko2026shearrate} derived a general expression for Taylor--Aris dispersion in steady, pressure-driven, shear-rate-dependent flows in uniform channels, allowing the transverse solute diffusivity to depend on viscosity through a generalised Stokes--Einstein relation. Their analysis distinguishes the effect of rheology on the velocity profile from its possible effect on molecular transport. The present formulation treats $D$ as a constant molecular diffusivity and does not account for viscosity-dependent diffusivity.

In this work, we extend the Taylor–Aris framework to the transport of a retained neutral solute in the pressure-driven flow of a power-law fluid (in contrast to \citep{gan2025antidispersion}) through a semipermeable circular tube. The coupled effects of fluid rheology and solvent filtration modify the axial evolution of the hydrodynamic field. This effect modifies solute dispersion and, depending on the conditions, can narrow or spread the solute band along the tube. Using multiple-scale homogenisation \citep{mei2010homogenization} of the convection–diffusion equation, we derive a one-dimensional macrotransport equation with spatially varying effective advection and dispersion coefficients. In addition, based on analysis of the axial variance, we determine the conditions under which the solute pulse narrows or spreads.

\section{Setup of the physical problem}

\begin{figure}[htbp]
    \centering
    \includegraphics[width=\linewidth]{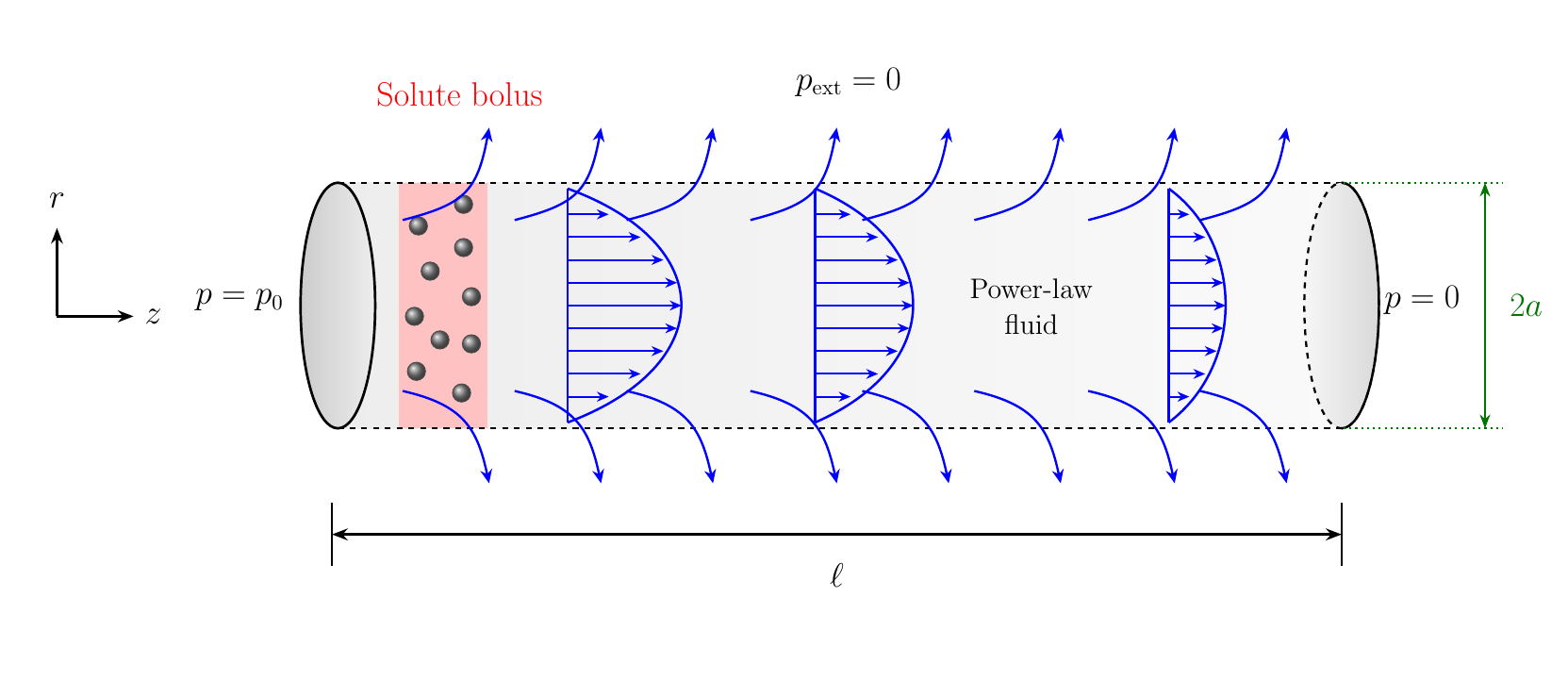}
    \captionsetup{skip=4pt}
\caption{Schematic diagram showing solute transport by the pressure-driven flow of a power-law fluid through a semipermeable circular tube of radius \(a\) and length \(\ell\). An initially localised solute bolus is advected downstream. The radial and axial coordinates are denoted by $r$ and $z$, respectively. The internal gauge pressure decreases from $p=p_0$ at the inlet to $p=0$ at the outlet, with the uniform external pressure taken as the reference ($p_{\mathrm{ext}}=0$). The blue internal curves illustrate the axially varying axial velocity profiles, while the outward blue arrows indicate solvent filtration through the tube wall. The wall has uniform hydraulic conductivity $\mathcal{L}_w$ and permits solvent permeation according to Starling's filtration law while preventing solute passage.}
    \label{Fig1}
\end{figure}

Consider the transport of a neutral solute by the steady,
axisymmetric flow of a power-law fluid through a slender circular
tube with $a\ll\ell$, as shown in figure~\ref{Fig1}. Here, $a$ and $\ell$ are the radius and length of the tube, respectively. The tube is rigid and has a semipermeable wall of constant hydraulic conductivity $\mathcal{L}_w$, which is defined as the wall-normal solvent velocity per unit transmural pressure difference (in units of $\mathrm{m\,(Pa\,s)^{-1}}$). The ratio $\epsilon=a/\ell\ll1$ is the tube aspect ratio.
The wall allows outward filtration of the solvent under a positive
transmural pressure difference preventing solute passage. This selective permeability corresponds to the ideal size-exclusion limit, in which membrane pores permit solvent permeation while excluding the solute species, as discussed by \citet{gan2025antidispersion}. Complete solute rejection is therefore imposed as a zero-total-normal-solute-flux condition at the wall.

We adopt the cylindrical coordinate system $(r,z)$, where $r$ and $z$ are the radial and axial coordinates, respectively. For axisymmetric
flow without swirl, the velocity field is expressed as
$\boldsymbol{u}=u_r(r,z)\,\boldsymbol{e}_r
+u_z(r,z)\,\boldsymbol{e}_z$, where $u_r$ and $u_z$ are the radial
and axial velocity components, and $\boldsymbol{e}_r$ and
$\boldsymbol{e}_z$ are the corresponding unit vectors.

At leading order in the slender-tube approximation, the pressure
depends only on $z$. Taking the uniform external pressure as the
reference, $p_{\mathrm{ext}}=0$, where $p_{\mathrm{ext}}$ is the external pressure, gives $p(z)$ as the internal gauge pressure and transmural pressure difference. The flow is driven
by an imposed pressure difference between the inlet and outlet,
with $p(0)=p_0$ and $p(\ell)=0$, where $p_0>0$ is the imposed inlet gauge pressure.

A dilute solute pulse, initially centred at $z=z_0$, is transported by the flow. Although the flow is steady, the solute concentration $c(r,z,t)$, where $t$ denotes time, evolves through advection and molecular diffusion. The interaction between radial variations in axial velocity and transverse molecular diffusion promotes axial dispersion. Meanwhile, solvent filtration through the wall
induces axial flow deceleration \citep{middleman1995modeling}, which can counteract spreading and increase the solute concentration \citep{gan2025antidispersion}.

\section{Governing equations}
\subsection{Hydrodynamic formulation}

For steady, incompressible, axisymmetric flow with negligible
inertial effects, the leading-order dimensional momentum and continuity equations in the lubrication limit are \citep{leal2007advanced}:
\begin{equation}
\frac{\partial p}{\partial r} \approx 0,
\qquad
-\frac{\mathrm{d}p}{\mathrm{d}z}
-\frac{1}{r}\frac{\partial}{\partial r}\left(r\tau_{rz}\right)=0,
\qquad
\frac{1}{r}\frac{\partial}{\partial r}(r u_r)
+\frac{\partial u_z}{\partial z}=0.
\label{eq:dimensional_lubrication_equations}
\end{equation}
Here, $p$ is the internal gauge pressure, and $\tau_{rz}$ is the positive shear stress, related to the shear component of the Cauchy stress tensor by $\tau_{rz}=-\sigma_{rz}$. The rheological response follows the power-law constitutive model, which at leading lubrication order is given by \citep{bird1987dynamics}:
\begin{equation}
\tau_{rz}
=-m\left|\frac{\partial u_z}{\partial r}\right|^{n-1}
\frac{\partial u_z}{\partial r}.
\label{eq:dimensional_power_law}
\end{equation}
Here, $m>0$ is the consistency index (in units of $\mathrm{Pa\,s^n}$) and $n>0$ is the flow behaviour index. The ranges $n<1$, $n=1$ and $n>1$ correspond to shear-thinning, Newtonian and shear-thickening responses, respectively. The numerical comparisons below consider $0<n\leq1$. For the pressure-driven flow considered here, $\mathrm{d}p/\mathrm{d}z < 0$ and
$\partial u_z/\partial r<0$ for $0<r\leq a$. Consequently,
$\tau_{rz}>0$ away from the axis, and the constitutive relation
can be written equivalently as:
\begin{equation}
\tau_{rz}=m\left(-\frac{\partial u_z}{\partial r}\right)^n.
\label{eq:dimensional_power_law_monotone}
\end{equation}

The velocity boundary conditions enforce symmetry about the tube's axis and no-slip at the wall. Solvent filtration is governed by Starling's filtration law \citep{starling1896absorption}, with osmotic pressure differences across the membrane neglected: 
\begin{equation}\label{BC1}
   \left.\frac{\partial u_z}{\partial r}\right|_{r=0}=0,
\qquad u_r(0,z)=0, 
\end{equation}
and
\begin{equation}
    u_z(a,z)=0,
\qquad
u_r(a,z)=\mathcal{L}_w[p(z)-p_{\mathrm{ext}}].
\end{equation}

The centreline condition~\eqref{BC1} also yields zero shear stress at $r=0$. Neglecting osmotic pressure is consistent with \citet{gan2025antidispersion}, provided its contribution is small relative to the hydrostatic transmural pressure difference in the region where the solute pulse is present.

\subsection{Dimensionless hydrodynamic formulation}
To express the governing equations that describe the physical problem in dimensionless form
and identify the parameters governing the flow, we introduce
the following dimensionless variables:
\begin{equation*}
R=\frac r a,\quad Z=\frac z\ell,\quad P=\frac p{p_0},\quad
U=\frac{u_z}{\mathcal V},\quad V=\frac{u_r}{\epsilon\mathcal V},\quad
\mathcal T_{RZ}=\frac{\tau_{rz}}{\mathcal T},
\end{equation*}
where
\begin{equation*}
\mathcal V=\left(\frac{p_0a^{n+1}}{m\ell}\right)^{1/n},\qquad
\mathcal T=m\left(\frac{\mathcal V}{a}\right)^n.
\end{equation*}
Here, $\mathcal V$ and $\mathcal T$ denote the characteristic
axial velocity and shear stress, respectively. The variables
$R$, $Z$ and $P$ denote the dimensionless radial coordinate,
axial coordinate and pressure, respectively; $U$ and $V$
are the dimensionless axial and radial velocity components;
and $\mathcal T_{RZ}$ is the dimensionless shear stress.

Introducing the dimensionless variables into the system \eqref{eq:dimensional_lubrication_equations}, we obtain
\begin{equation}
\frac{\partial P}{\partial R} = 0, \qquad -\frac{\mathrm{d}P}{\mathrm{d}Z} - \frac{1}{R}\frac{\partial}{\partial R}\left( R \mathcal{T}_{RZ} \right) = 0, \qquad \frac{1}{R}\frac{\partial}{\partial R}(RV) + \frac{\partial U}{\partial Z} = 0.
\label{eq:dimensionless_lubrication_equations}
\end{equation}
The dimensionless version of the constitutive equation \eqref{eq:dimensional_power_law_monotone} is: 
\begin{equation}
\mathcal{T}_{RZ}
=\left(-\frac{\partial U}{\partial R}\right)^n.
\label{eq:dimensionless_power_law}
\end{equation}

To derive the hydrodynamic field, we first integrate the dimensionless axial momentum equation and apply the symmetry condition at $R=0$. This yields the following result:
\begin{equation}
\mathcal{T}_{RZ}=-\frac{R}{2}\frac{\mathrm{d}P}{\mathrm{d}Z},
\label{eq:dimensionless_shear_stress}
\end{equation}
 Combining this result with Eq.~\eqref{eq:dimensionless_power_law} yields
\begin{equation}
\frac{\partial U}{\partial R}
=-\left(-\frac{R}{2}\frac{\mathrm{d}P}{\mathrm{d}Z}\right)^{1/n}.
\label{eq:dimensionless_velocity_gradient}
\end{equation}
Integrating Eq. \eqref{eq:dimensionless_velocity_gradient} and applying the no-slip boundary condition, $U(1,Z)=0$, yields:
\begin{equation}
U(R,Z) = B(Z)\left(1 - R^{1+1/n}\right), \quad \text{where} \quad B(Z) = \frac{n}{n+1}\left(-\frac{1}{2}\frac{\mathrm{d}P}{\mathrm{d}Z}\right)^{1/n}.
\label{eq:axial_velocity_B}
\end{equation}
Here $B(Z)=U(0,Z)$ is the dimensionless centreline axial velocity. Substituting the axial velocity profile \eqref{eq:axial_velocity_B} into the continuity equation in \eqref{eq:dimensionless_lubrication_equations}, integrating with respect to $R$, and applying
the axis boundary condition $V(0,Z)=0$, which corresponds to the
second condition in \eqref{BC1}, yields
\begin{equation}
V(R,Z) = -\frac{\mathrm{d}B}{\mathrm{d}Z}\left[ \frac{R}{2} - \frac{R^{2+1/n}}{3+1/n} \right].
\label{eq:radial_velocity}
\end{equation}

At $R=1$, Starling's law gives $V(1,Z)=\Lambda P(Z)$, with
\begin{equation*}
\Lambda=\frac{\mathcal L_w p_0}{\epsilon\mathcal V}
=\frac{\mathcal{L}_w m^{1/n}p_0^{1-1/n}\ell^{1+1/n}}{a^{2+1/n}}.
\end{equation*}
The filtration parameter $\Lambda$ measures the ratio of the characteristic wall-normal solvent velocity to the radial velocity scale $\epsilon\mathcal V$. It therefore incorporates the effects of membrane hydraulic conductivity, imposed pressure, tube geometry and power-law rheology.

When the radial velocity \eqref{eq:radial_velocity} is evaluated at the wall ($R=1$), $B(Z)$ is replaced using \eqref{eq:axial_velocity_B}, and the filtration boundary condition $V(1,Z)=\Lambda P(Z)$ is imposed, the result is the following nonlinear differential equation for the pressure, subject to the prescribed inlet
and outlet pressures:
\begin{equation}
-\frac{n}{2(3n+1)}\frac{\mathrm{d}}{\mathrm{d}Z}\left[ \left( -\frac{1}{2}\frac{\mathrm{d}P}{\mathrm{d}Z} \right)^{1/n} \right] = \Lambda P(Z), \quad \text{with} \quad P(0)=1, \; P(1)=0.
\label{eq:pressure_equation}
\end{equation}

The averaging operation is $\langle f \rangle = 2\int_0^1 R f(R,Z) \, \mathrm{d}R$, where $\langle f\rangle$ is the cross-sectional average of an axisymmetric field $f$. It yields
\begin{equation}
\langle U\rangle(Z)=\frac{n}{3n+1}
\left(-\frac{1}{2}\frac{\mathrm{d}P}{\mathrm{d}Z}\right)^{1/n},
\label{eq:mean_velocity}
\end{equation}
where $\langle U\rangle(Z)$ is the dimensionless mean axial velocity. Averaging the continuity equation over the cross-section yields the global mass balance:
\begin{equation}
\frac{\mathrm{d}\langle U \rangle}{\mathrm{d}Z} = -2V(1,Z) = -2\Lambda P(Z).
\label{eq:mean_velocity_gradient}
\end{equation}
This relation, given by \eqref{eq:mean_velocity_gradient}, explicitly links solvent loss through the wall to the downstream decrease in mean axial velocity. The resulting axial deceleration forms the hydrodynamic basis for solute compression, which can counteract dispersive spreading.

Our analysis is restricted to parameter regimes in which a non-zero
proportion of the incoming fluid reaches the outlet; that is,
$\langle U\rangle(1)>0$, or equivalently, $\left.\frac{\mathrm{d}P}{\mathrm{d}Z}\right|_{Z=1}<0$.
By integrating \eqref{eq:mean_velocity_gradient}, this requirement implies that:
\begin{equation}
    2\Lambda\int_0^1 P(Z)\,\mathrm{d}Z<\langle U\rangle(0),
\end{equation}
which states that the total fluid loss through the wall must remain smaller than the inlet flow rate. This inequality limits the analysis to solutions with positive outlet throughflow and is not an additional boundary condition.

The limiting state $\left.\frac{\mathrm{d}P}{\mathrm{d}Z}\right|_{Z=1}=0$ determines the critical filtration parameter. Appendix~\ref{app:critical_filtration} derives this hydraulic limit from a first integral of the pressure equation and explains why it is finite only for $0<n<1$. It gives
\begin{equation}
\Lambda_c(n)
=
\frac{2n}{(n+1)(3n+1)}
\left[
\frac{n+1}{2(1-n)}
\right]^{(n+1)/n},
\qquad 0<n<1.
\label{eq:critical_filtration}
\end{equation}
Positive outlet throughflow therefore requires $\Lambda<\Lambda_c(n)$, while $\Lambda=\Lambda_c(n)$ marks the boundary of the positive-outlet-throughflow regime. 

In the Newtonian case ($n=1$), the pressure equation reduces to $\frac{\mathrm{d}^2P}{\mathrm{d}Z^2}=16\Lambda P$.
For $\Lambda>0$, its solution is
\begin{equation}
P(Z)=\frac{\sinh\!\bigl[4\sqrt{\Lambda}(1-Z)\bigr]}{\sinh\!\bigl(4\sqrt{\Lambda}\bigr)},
\qquad
\langle U\rangle(Z)=\frac{\sqrt{\Lambda}}{2}
\frac{\cosh\!\bigl[4\sqrt{\Lambda}(1-Z)\bigr]}{\sinh\!\bigl(4\sqrt{\Lambda}\bigr)}.
\label{eq:newtonian_hydrodynamics}
\end{equation}
For an impermeable wall, the limit $\Lambda\to0$ recovers the hydrodynamics results of the well-known Poiseuille flow \cite{middleman1995modeling}, $P=1-Z$ and $\langle U\rangle=1/8$.

\section{Convection--diffusion equation}

We now consider the transport of a dilute solute by the steady
velocity field obtained above. The solute is assumed to be
passive, with constant molecular diffusivity $D$, and is fully
retained by the semipermeable wall, through which solvent
filtration occurs. Under axisymmetric conditions, the
concentration $c(r,z,t)$ evolves according to
\begin{equation}
\frac{\partial c}{\partial t}
+ u_z \frac{\partial c}{\partial z}
+ u_r \frac{\partial c}{\partial r}
= D \left[
\frac{\partial^2 c}{\partial z^2}
+ \frac{1}{r}\frac{\partial}{\partial r}
\left(r\frac{\partial c}{\partial r}\right)
\right].
\end{equation}
Symmetry at the tube axis requires
\begin{equation}
\left.\frac{\partial c}{\partial r}\right|_{r=0}=0,
\qquad 0<z<\ell,\quad t>0.
\label{eq:dimensional_solute_axis_condition}
\end{equation}
At the semipermeable membrane, achieving complete solute retention requires that the net normal solute flux to vanish:
\begin{equation}
\left.\left(u_r c-D\frac{\partial c}{\partial r}\right)\right|_{r=a}=0,
\qquad 0<z<\ell,\quad t>0.
\label{eq:dimensional_solute_wall_flux}
\end{equation}
This condition can be expressed using Starling's filtration law \citep{starling1896absorption}:
\begin{equation}
D\left.\frac{\partial c}{\partial r}\right|_{r=a}
=\mathcal L_w[p(z)-p_{\mathrm{ext}}]c(a,z,t).
\label{eq:dimensional_solute_wall_condition}
\end{equation}
During outward solvent filtration, the inward diffusive flux of solute
therefore balances the outward advective solute flux.
As a result, achieving complete solute retention typically requires
a nonzero radial concentration gradient at the membrane wall.

For the finite-domain pulse calculations, the dimensional axial boundary conditions are
\begin{equation}
c(r,0,t)=0,
\qquad
\left.\frac{\partial c}{\partial z}\right|_{z=\ell}=0,
\qquad 0\leq r\leq a,\quad t>0.
\label{eq:dimensional_solute_axial_conditions}
\end{equation}
These conditions specify that, after pulse initialisation, the inlet solute concentration is zero, and a zero axial diffusive flux is enforced at the outlet, allowing solute to exit only by advection.
The solute pulse is initialised within the tube, rather than being supplied continuously at the inlet. Its initial dimensional concentration is prescribed as
\begin{equation}
c(r,z,0)=c_0\exp\!\left[-\frac{(z-z_0)^2}{2s_0^2}\right],
\qquad 0\leq r\leq a,\quad 0<z<\ell,
\label{eq:dimensional_initial_pulse}
\end{equation}
where $c_0$ is the initial peak concentration, $z_0$ is the axial location of the pulse centre, and $s_0$ is the Gaussian axial standard-deviation parameter. The pulse is initially uniform across each cross-section. We choose $z_0$ sufficiently far from both ends relative to $s_0$ so the boundary tails are negligible. Under outward filtration, the initially uniform radial distribution does not exactly satisfy the wall condition~\eqref{eq:dimensional_solute_wall_condition} and develops a radial concentration gradient during an initial diffusive adjustment. The axial boundary conditions apply for $t>0$; the Gaussian tails need not satisfy them exactly at $t=0$.

The convection-diffusion equation is nondimensionalised using the following dimensionless variables:
\begin{equation*}
C=\frac c{c_0},\qquad T=\frac {t}{a^2/D},
\end{equation*}
 where $C$ is the dimensionless concentration and $T$ is the dimensionless time. We also define the dimensionless parameters $Z_0=z_0/\ell$ and $\sigma_0=s_0/\ell$. Defining the diffusive Peclet number $Pe=\mathcal Va/D$ and using $\epsilon=a/\ell$, the transport equation becomes
\begin{equation}
\frac{\partial C}{\partial T} + \epsilon Pe \left( U \frac{\partial C}{\partial Z} + V \frac{\partial C}{\partial R} \right) = \epsilon^2 \frac{\partial^2 C}{\partial Z^2} + \frac{1}{R}\frac{\partial}{\partial R}\left( R \frac{\partial C}{\partial R} \right).
\label{eq:dimensionless_convection_diffusion}
\end{equation}

The centreline boundary condition requires axisymmetry, $\left.\partial C / \partial R \right|_{R=0} = 0$. At the semipermeable wall ($R=1$), solute retention requires zero net normal solute flux $J_R = \epsilon Pe\,V C - \partial C/\partial R = 0$, where $J_R$ is scaled by $Dc_0/a$. Expressing this boundary condition in terms of filtration velocity yields
\begin{equation}
\left. \frac{\partial C}{\partial R} \right|_{R=1} = \epsilon Pe \, V(1,Z) C(1,Z,T) = \epsilon Pe \, \Lambda P(Z) C(1,Z,T).
\label{eq:dimensionless_solute_wall_condition}
\end{equation}

\section{Multiple-scale formulation and Taylor dispersion}
\label{sec:multiple_scales}

After the initial transverse diffusion transient, the solute concentration evolves over longer time scales associated with axial advection and dispersion. This separation of scales reduces the cross-sectional transport problem to a one-dimensional equation for the mean concentration while retaining the effects of the non-uniform velocity field and solvent filtration. To derive this post-transient macrotransport equation, we use the multiple-scale homogenisation method described by \citet{mei2010homogenization} and introduce the hierarchy of time scales
\begin{equation}
T_0 = T, \qquad T_1 = \epsilon T, \qquad T_2 = \epsilon^2 T.
\label{eq:time_scales}
\end{equation}
Here, $T_0$, $T_1$, and $T_2$ denote the fast, intermediate and slow time variables, respectively. This hierarchy transforms the temporal partial derivative into
\begin{equation}
\frac{\partial}{\partial T} = \frac{\partial}{\partial T_0} + \epsilon \frac{\partial}{\partial T_1} + \epsilon^2 \frac{\partial}{\partial T_2}.
\label{eq:time_derivative_expansion}
\end{equation}
The concentration field is expanded as a power series in the slender-tube parameter, $\epsilon$:
\begin{equation}
C(R,Z,T_0,T_1,T_2) = C_0 + \epsilon C_1 + \epsilon^2 C_2 + O(\epsilon^3).
\label{eq:concentration_expansion}
\end{equation}
Here $C_0$ is the leading-order concentration, and $C_1$ and $C_2$ are its first- and second-order corrections. The expansion assumes that the hydrodynamic parameters and $Pe$ are fixed as $\epsilon\to0$. The first-order response increases with $Pe$ and axial concentration gradients, meaning that transverse equilibration is particularly important for narrow pulses and strong filtration.

\subsection{Asymptotic expansion at lower orders}

At $O(1)$, the transport problem results in rapid radial diffusion across the cross-section of the tube:
\begin{equation}
\frac{\partial C_0}{\partial T_0} = \frac{1}{R} \frac{\partial}{\partial R} \left( R \frac{\partial C_0}{\partial R} \right).
\label{eq:order_one_transport}
\end{equation}
Once the initial cross-sectional diffusive transient has decayed ($T_0 \to \infty$), the leading-order concentration becomes radially uniform:
\begin{equation}
C_0 = C_0(Z,T_1,T_2).
\label{eq:C0_uniform}
\end{equation}

At $O(\epsilon)$, the governing perturbation equation takes the form
\begin{equation}
\frac{\partial C_1}{\partial T_0} + \frac{\partial C_0}{\partial T_1} + Pe \, U(R,Z)\frac{\partial C_0}{\partial Z} = \frac{1}{R}\frac{\partial}{\partial R}\left( R \frac{\partial C_1}{\partial R} \right),
\label{eq:order_epsilon}
\end{equation}
subject to zero radial solute flux at the centreline,
\begin{equation}
\left. \frac{\partial C_1}{\partial R} \right|_{R=0} = 0.
\label{eq:order_epsilon_bc_centreline}
\end{equation}
At the semipermeable wall, the first-order boundary condition becomes
\begin{equation}
\left. \frac{\partial C_1}{\partial R} \right|_{R=1} = Pe \, V(1,Z) C_0.
\label{eq:order_epsilon_bc_wall}
\end{equation}

Averaging equation~\eqref{eq:order_epsilon} across the cross-section, and applying the normalisation $\langle C_1\rangle=0$, yields the first-order solvability condition
\begin{equation}
\frac{\partial C_0}{\partial T_1} + Pe \, \frac{\partial}{\partial Z}\left( \langle U \rangle C_0 \right) = 0,
\label{eq:order_epsilon_macroscopic}
\end{equation}
which describes advection governed by the cross-sectionally averaged axial velocity.

\subsection{First-order cell problems and explicit solutions}

By subtracting the solvability condition \eqref{eq:order_epsilon_macroscopic} from the $O(\epsilon)$ balance \eqref{eq:order_epsilon}, once the radial transient has decayed ($\partial C_1/\partial T_0 = 0$), one obtains the steady radial fluctuation problem:
\begin{equation}
\frac{1}{R}\frac{\partial}{\partial R}\left( R \frac{\partial C_1}{\partial R} \right) = Pe (U - \langle U \rangle) \frac{\partial C_0}{\partial Z} + 2 Pe \, V(1,Z) C_0.
\label{eq:C1_steady_radial_problem}
\end{equation}

By linear superposition, the first-order correction to the concentration can be decomposed into zero-mean components arising from axial velocity variations and wall filtration:
\begin{equation}
C_1 = Pe \, \chi(R,Z) \frac{\partial C_0}{\partial Z} + Pe \, \psi(R,Z) C_0,
\label{eq:C1_ansatz}
\end{equation}
where $\chi(R,Z)$ and $\psi(R,Z)$ are the dimensionless radial response functions for shear dispersion and wall filtration, respectively, each subject to zero-mean normalisations $\langle \chi \rangle = 0$ and $\langle \psi \rangle = 0$. The normalisation $\langle C_j\rangle = 0$, with $j = 1, 2$, is imposed on the higher-order corrections, ensuring that $C_0$ denotes the cross-sectional mean to the order considered. The cell problems characterise the post-transient response; an initially radially uniform Gaussian does not precisely satisfy the permeable-wall concentration condition, resulting in an initial radial adjustment.

The shear-dispersion cell function, $\chi(R,Z)$, quantifies the contribution of radial variations in the axial velocity, given by $U - \langle U \rangle = B [ 2/(q+2) - R^q ]$, where $q = 1 + 1/n$:
\begin{equation}
\frac{1}{R}\frac{\partial}{\partial R}\left( R \frac{\partial \chi}{\partial R} \right) = U - \langle U \rangle, \quad \left.\frac{\partial\chi}{\partial R}\right|_{R=0,1} = 0.
\end{equation}
The following explicit solution is obtained by integrating twice:
\begin{equation}
\chi(R,Z) = \frac{B(Z)}{q+2} \left[ \frac{R^2}{2} - \frac{R^{q+2}}{q+2} - \frac{1}{4} + \frac{2}{(q+2)(q+4)} \right].
\label{eq:chi_solution}
\end{equation}

The wall-filtration cell function, $\psi(R,Z)$, characterises the redistribution of solute arising from solvent extraction at the boundary:
\begin{equation}
\frac{1}{R}\frac{\partial}{\partial R}\left( R \frac{\partial \psi}{\partial R} \right) = 2V(1,Z), \quad \left.\frac{\partial\psi}{\partial R}\right|_{R=0} = 0, \quad \left.\frac{\partial\psi}{\partial R}\right|_{R=1} = V(1,Z).
\end{equation}
Direct integration gives
\begin{equation}
\psi(R,Z) = \frac{V(1,Z)}{2}\left( R^2 - \frac{1}{2} \right) = \frac{\Lambda P(Z)}{2}\left( R^2 - \frac{1}{2} \right).
\label{eq:psi_solution}
\end{equation}

\subsection{Second-order macrotransport and effective coefficients}

At $O(\epsilon^2)$, the concentration perturbation satisfies
\begin{equation}
\frac{\partial C_2}{\partial T_0} + \frac{\partial C_1}{\partial T_1} + \frac{\partial C_0}{\partial T_2} + Pe \, U \frac{\partial C_1}{\partial Z} + Pe \, V \frac{\partial C_1}{\partial R} = \frac{\partial^2 C_0}{\partial Z^2} + \frac{1}{R}\frac{\partial}{\partial R}\left( R \frac{\partial C_2}{\partial R} \right).
\end{equation}

To obtain the averaged second-order equation, continuity gives
\begin{equation*}
\left\langle U\frac{\partial C_1}{\partial Z}+V\frac{\partial C_1}{\partial R}\right\rangle
=\frac{\partial}{\partial Z}\langle U C_1\rangle+2V(1,Z)C_1(1).
\end{equation*}
The radial diffusion term averages to $2\left.\frac{\partial C_2}{\partial R}\right|_{R=1}=2PeV(1,Z)C_1(1)$. The wall terms cancel, while the normalised corrections have zero averaged time derivatives after the radial transient. Substituting equation~\eqref{eq:C1_ansatz} then gives
\begin{equation}
\frac{\partial C_0}{\partial T_2} = \frac{\partial^2 C_0}{\partial Z^2} - Pe^2 \frac{\partial}{\partial Z}\left[ \langle U\chi \rangle \frac{\partial C_0}{\partial Z} + \langle U\psi \rangle C_0 \right].
\label{eq:second_order_effective_general}
\end{equation}

Using~\eqref{eq:chi_solution} and~\eqref{eq:psi_solution} to evaluate the cross-sectional averages gives:
\begin{equation}
\langle U\chi \rangle = -\frac{n^2 \langle U \rangle^2}{2(3n+1)(5n+1)}, \quad \langle U\psi \rangle = -\frac{n^2 \Lambda P(Z)}{2(3n+1)(5n+1)} \left(-\frac{1}{2}\frac{\mathrm{d}P}{\mathrm{d}Z}\right)^{1/n}.
\end{equation}

By combining the solvability conditions at $O(\epsilon)$ and $O(\epsilon^2)$, one obtains the one-dimensional macrotransport equation, accurate to the retained asymptotic order:
\begin{equation}
\frac{\partial C_0}{\partial T} + \frac{\partial}{\partial Z} \left[ \mathcal{A}_{\mathrm{eff}}(Z) C_0 \right] = \epsilon^2 \frac{\partial}{\partial Z} \left[ D_{\mathrm{eff}}(Z) \frac{\partial C_0}{\partial Z} \right] + O(\epsilon^3),
\label{eq:macro_transport_advective_form}
\end{equation}
where $\mathcal{A}_{\mathrm{eff}}(Z)$ denotes the dimensionless effective axial advection coefficient, and $D_{\mathrm{eff}}(Z)$ is the total effective axial dispersion coefficient, normalised by $D$. The effective advection coefficient satisfies
\begin{equation}
\mathcal{A}_{\mathrm{eff}}(Z) = \epsilon Pe \, \langle U \rangle(Z) - \epsilon^2 \mathcal{K}_w(Z).
\label{eq:effective_advection}
\end{equation}
Here, $\mathcal{K}_w(Z)$ denotes the dimensionless wall-filtration correction to the effective advection. As $\langle U \psi \rangle < 0$ under outward filtration, this correction reduces the effective advection coefficient:
\begin{equation*}
\mathcal{K}_w(Z) = -Pe^2 \langle U \psi \rangle\geq0.
\end{equation*}

Evaluating the dispersion coefficient gives
\begin{equation}
D_{\mathrm{eff}}(Z)=1+D_{\mathrm{T}}(Z),\qquad
D_{\mathrm{T}}(Z)=-Pe^2\langle U\chi\rangle
=Pe^2\frac{n^2\langle U\rangle(Z)^2}{2(3n+1)(5n+1)},
\label{eq:composite_Deff}
\end{equation}
where $D_{\mathrm T}(Z)$ is the dimensionless Taylor contribution, and the unit term represents molecular diffusion. For $n>0$, the above expression gives $D_{\mathrm T}\geq0$, so $D_{\mathrm{eff}}\geq1$ throughout the model. The wall filtration coefficient is
\begin{equation}
\mathcal{K}_w(Z) = Pe^2 \frac{n^2 \Lambda P(Z)}{2(3n+1)(5n+1)} \left( -\frac{1}{2}\frac{\mathrm{d}P}{\mathrm{d}Z} \right)^{1/n}.
\label{eq:composite_Kw}
\end{equation}

\subsection{Particular cases recovered from the derived solution}
\label{sec:limiting_cases}

The derived hydrodynamic fields and transport coefficients recover a range of limiting and particular cases. The corresponding concentration profiles satisfy the macrotransport equation to the retained asymptotic order; exact finite-domain solutions are distinguished from infinite-domain approximations, in which axial boundary effects are neglected.

\noindent\textit{Impermeable wall, $\Lambda=0$.}
The pressure and mean velocity become
\begin{equation}
P=1-Z,\qquad u_*\equiv\langle U\rangle=\frac{n}{3n+1}2^{-1/n},\qquad
\mathcal K_w=0.
\label{eq:impermeable_limit}
\end{equation}
Here, $u_*$ denotes the constant mean axial velocity of the impermeable tube. The quantities $A_* = \epsilon Pe\,u_*$ and $D_* = 1 + Pe^2 f(n) u_*^2$ are the constant effective advection and dispersion coefficients, respectively, where $f(n) = n^2 / [2 (3n + 1)(5n + 1)]$ is the rheology-dependent velocity-profile factor. For an initial Gaussian pulse, the infinite-domain solution is
\begin{equation}
C_0(Z,T)=\frac{\sigma_0}{s(T)}
\exp\left[-\frac{(Z-Z_0-A_*T)^2}{2s(T)^2}\right],\qquad
s(T)^2=\sigma_0^2+2\epsilon^2D_*T.
\label{eq:impermeable_gaussian_limit}
\end{equation}
Here, $s(T)$ denotes the axial standard deviation of the infinite-domain Gaussian, with $s(0) = \sigma_0$. This serves as a reference approximation for the finite tube in the limit where axial boundary effects are negligible. The solution exhibits monotonic broadening and does not account for filtration-driven antidispersion. It further provides an analytical benchmark for pulse position, width, and amplitude.

\noindent\textit{Newtonian fluid, $n=1$.}
Together with the exact hydrodynamic solution~\eqref{eq:newtonian_hydrodynamics}, the coefficients reduce to
\begin{equation}
D_{\mathrm{eff}}=1+\frac{Pe^2\langle U\rangle^2}{48},\qquad
\mathcal K_w=\frac{Pe^2\Lambda P\langle U\rangle}{12}.
\label{eq:newtonian_transport_limit}
\end{equation}
For a permeable Newtonian tube, we define $k = 4\sqrt{\Lambda}$ and $x = k(1 - Z)$, where $k$ denotes the dimensionless pressure-decay parameter and $x$ is the scaled axial distance from the outlet. Substituting \eqref{eq:newtonian_hydrodynamics} into \eqref{eq:newtonian_transport_limit}, together with \eqref{eq:effective_advection}, yields
\begin{equation}
\begin{aligned}
D_{\mathrm{eff}}(Z)
&=1+\frac{Pe^2\Lambda}{192}
\frac{\cosh^2 x}{\sinh^2 k},\\[4pt]
\mathcal K_w(Z)
&=\frac{Pe^2\Lambda^{3/2}}{24}
\frac{\sinh x\cosh x}{\sinh^2 k},\\[4pt]
\mathcal A_{\mathrm{eff}}(Z)
&=\frac{\epsilon Pe\sqrt{\Lambda}}{2}
\frac{\cosh x}{\sinh k}
-\frac{\epsilon^2 Pe^2\Lambda
P(Z)\langle U\rangle(Z)}{12}.
\end{aligned}
\label{eq:newtonian_permeable_explicit_case}
\end{equation}

For $\Lambda > 0$, $D_{\mathrm{eff}}$ decreases along the length of the tube, while $\mathcal{K}_w(Z)$ vanishes at the outlet. Nevertheless, $D_{\mathrm{eff}}(1) = 1 + Pe^2 \Lambda / (192 \sinh^2 k) > 1$ for $Pe > 0$, consistent with finite outlet flow. Expanding the hyperbolic functions as $\Lambda \to 0$ removes the apparent singularities and recovers the impermeable Newtonian limit \citep{taylor1953dispersion}. For $\Lambda > 0$, axial variation in the effective advection and dispersion coefficients precludes an exact Gaussian solution of the type obtained for constant coefficients.

\noindent\textit{Classical Poiseuille--Taylor transport, $n=1$ and $\Lambda=0$.}
Taking both limits yields the following limiting expressions
\begin{equation}
P=1-Z,\quad U=\frac14(1-R^2),\quad V=0,\quad
\mathcal A_{\mathrm{eff}}=\frac{\epsilon Pe}{8},\quad
D_{\mathrm{eff}}=1+\frac{Pe^2}{3072}.
\label{eq:poiseuille_taylor_case}
\end{equation}
Here, $Pe=\mathcal V a/D$ uses the characteristic velocity $\mathcal V$, whereas the dimensional mean velocity in this case is $\overline u_z=\mathcal V/8$. The quantity $Pe_m=\overline u_z a/D=Pe/8$ is the radius-based mean-flow P\'eclet number. It gives
\begin{equation}
D_{\mathrm{eff}}=1+\frac{Pe_m^2}{48},\qquad
D_{\mathrm{eff}}^{\mathrm{dim}}=D D_{\mathrm{eff}}
=D+\frac{\overline u_z^{\,2}a^2}{48D}.
\label{eq:classical_mean_velocity_scaling}
\end{equation}
Here, $D_{\mathrm{eff}}^{\mathrm{dim}}$ is the dimensional effective axial dispersion coefficient, with units of $\mathrm{m^2\,s^{-1}}$. Equation~\eqref{eq:poiseuille_taylor_case} is therefore identical to the classical Taylor--Aris result, with the distinct numerical denominator reflecting the velocity scale employed in $Pe$. If the tube diameter is used in place of the radius, $Pe_d = 2Pe_m$, and the corresponding result is $D_{\mathrm{eff}} = 1 + Pe_d^2 / 192$.

The corresponding Gaussian evolution can be derived directly from~\eqref{eq:impermeable_gaussian_limit}, which recovers the classical centroid, variance, and amplitude before axial boundary effects become significant.

\noindent\textit{Weak filtration, $\Lambda\to0^+$ at fixed $n>0$.}
Expanding equations~\eqref{eq:pressure_equation} and \eqref{eq:mean_velocity_gradient} about~\eqref{eq:impermeable_limit} gives
\begin{equation}
\begin{aligned}
\langle U\rangle(Z)&=u_*+\Lambda\left(Z^2-2Z+\frac23\right)+O(\Lambda^2),\\
P(Z)&=1-Z-\frac{n\Lambda}{3u_*}Z(Z-1)(Z-2)+O(\Lambda^2).
\end{aligned}
\label{eq:weak_filtration_limit}
\end{equation}
The constant $2/3$ arises from the imposed
pressure drop, indicating that filtration alters
both the inlet velocity and its downstream gradient.
The expansion remains valid provided the corrections are small relative to the leading-order terms; in particular, $\Lambda/u_* \ll 1$ is required for fixed $n$ of order unity. The corresponding transport coefficients are then given by
\begin{equation}
\begin{aligned}
D_{\mathrm{eff}}&=1+Pe^2f(n)u_*^2
+2\Lambda Pe^2f(n)u_*\left(Z^2-2Z+\frac23\right)+O(\Lambda^2),\\
\mathcal K_w&=\Lambda\frac{Pe^2n u_*}{2(5n+1)}(1-Z)+O(\Lambda^2).
\end{aligned}
\label{eq:weak_filtration_coefficients}
\end{equation}
These expressions provide the leading-order corrections attributable to weak filtration, without implying that such filtration necessarily contracts a pulse.

\noindent\textit{Outlet and critical filtration limits.}
Since $P(1) = 0$, every regular solution with a positive axial volumetric flow rate at the outlet satisfies $\mathcal{K}_w(1) = 0$, whereas
\begin{equation}
D_{\mathrm{eff}}(1)=1+Pe^2f(n)\langle U\rangle(1)^2>1
\quad\text{for }Pe>0.
\label{eq:outlet_transport_limit}
\end{equation}
Thus, vanishing wall filtration at the outlet does not, by itself, eliminate shear dispersion. For $0 < n < 1$, the limiting hydrodynamic state as $\Lambda \to \Lambda_c^-$ is instead characterised by $\langle U \rangle(1) \to 0$ and $D_{\mathrm{eff}}(1) \to 1$. At the critical state, the first integral in appendix~\ref{app:critical_filtration} yields
\begin{equation}
P_c(Z)=(1-Z)^{(n+1)/(1-n)},\qquad
\langle U\rangle_c(Z)=\frac{n}{3n+1}
\left[\frac{n+1}{2(1-n)}\right]^{1/n}
(1-Z)^{2/(1-n)}.
\label{eq:critical_profiles_limit}
\end{equation}
Here, the subscript $c$ denotes evaluation at the critical filtration state $\Lambda = \Lambda_c$.
These critical profiles describe the limiting hydraulic state, rather than a contraction threshold or an explicit pulse evolution. The accuracy of the transport reduction in the vicinity of this limit requires separate assessment.

\FloatBarrier
\subsection{Dimensional reference case}
\suppressfloats[t]
\label{sec:dimensional_reference}

Table~\ref{tab:dimensional_reference} summarises the common geometric scales and the distinct input parameters for the two cases considered. The Newtonian column translates the validation-case data of \citet{gan2025antidispersion} (see Appendix~\ref{app:variance_balance}) to a circular tube of radius equal to their planar half-width. The power-law column represents the chosen numerical reference, employing the $100$-nm tracer diffusivity and xanthan parameters from \citet{chakraborty2019powerlaw}.
\begin{table}
\centering
\small
\begin{tabular}{lll}
\hline
Parameter & Newtonian tube: Gan inputs & Power-law reference\\
\hline
$a=h$ & \multicolumn{2}{c}{$10\,\mu\mathrm m$ (common)}\\
$\ell=L$ & \multicolumn{2}{c}{$5\,\mathrm{mm}$ (common)}\\
$\epsilon=a/\ell$ & \multicolumn{2}{c}{$0.002$ (common)}\\
\hline
$n$ & $1$ & $0.4$\\
$\mu\ (\mathrm{Pa\,s})$ & $7.0\times10^{-4}$ & ---\\
$m\ (\mathrm{Pa\,s^n})$ & --- & $0.3592$\\
$D\ (\mathrm{m^2\,s^{-1}})$ & $1.0\times10^{-10}$ & $5.6\times10^{-12}$\\
$p_0\ (\mathrm{Pa})$ & $10$ & $163$\\
$\mathcal L_w\ (\mathrm{m\,(Pa\,s)^{-1}})$ & $2.86\times10^{-7}$ & $1.95\times10^{-11}$\\
\hline
$\mathcal V\ (\mu\mathrm{m\,s^{-1}})$ & $285.71$ & $7.84$\\
$Pe=\mathcal Va/D$ & $28.571$ & $14$\\
$\Lambda=\mathcal L_wp_0/(\epsilon\mathcal V)$ & $5.005$ & $0.2025$\\
\hline
\end{tabular}
\caption{Dimensional reference conditions for Newtonian
and power-law fluids. Here, $\mu$ denotes the Newtonian
dynamic viscosity, and $m$ is the power-law consistency
index. The constitutive laws coincide when $n = 1$ and
$m = \mu$. The Newtonian column uses the physical
parameters reported by \citet{gan2025antidispersion},
with the viscosity of water evaluated at
$37\,{}^\circ\mathrm{C}$; $\mathcal{V}$, $Pe$, and
$\Lambda$ are recalculated for the circular tube.
For the power-law fluid, the prescribed values of
$Pe$ and $\Lambda$, together with the selected
geometric, rheological, and molecular transport
parameters, determine $\mathcal{V}$, $p_0$, and
$\mathcal{L}_w$. The two columns correspond to
distinct operating conditions and tracer
diffusivities.}
\label{tab:dimensional_reference}\label{tab:gan_physical_conversion}
\end{table}

\citet{gan2025antidispersion} define $u_0 = p_0 h^2 / (2 \mu L)$ and $B_{\mathrm{G}}^2 = 3 L_p \mu L^2 / h^3$. Setting $a = h$, $\ell = L$, $m = \mu$, and $\mathcal{L}_w = L_p$, our Newtonian convention yields
\begin{equation}
\mathcal V=2u_0,\qquad Pe=2Pe_{\mathrm G},\qquad\Lambda=B_{\mathrm G}^2/3.
\label{eq:gan_physical_conversion}
\end{equation}
These relations map the reported values $Pe_{\mathrm{G}} = 14.3$ and $B_{\mathrm{G}} = 3.87$ to the Newtonian column. The chosen power-law reference corresponds to a distinct operating condition, with the dimensional pressure and conductivity determined from
\begin{equation}
p_0=m\,\frac{\ell}{a^{n+1}}\mathcal V^n,\qquad
\mathcal L_w=\frac{\Lambda\epsilon\mathcal V}{p_0}.
\label{eq:dimensional_reference_parameters}
\end{equation}
For $n=0.4$, the critical filtration parameter is
$\Lambda_c(0.4)\simeq0.4455$, so the reference case
satisfies the restriction $\Lambda\leq0.5\Lambda_c(n)$
used in the performance maps. The xanthan data exhibit
a low-shear Newtonian plateau. The pure power-law model
therefore provides an illustrative constitutive
approximation whose limitations are particularly
relevant near the axis and in other regions of low shear.

\FloatBarrier
\subsection{Reference pulse and operating conditions}
\label{sec:computational_description}
The reference pulse is characterised by $Pe = 14$, $\epsilon = 0.002$, $Z_0 = 0.25$, and $\sigma_0 = 0.06$. Comparisons between $n = 1$ and $n = 0.4$, with $\Lambda = 0$ or $0.2025$, elucidate the influence of rheology and solvent filtration on pulse displacement and deformation. Cross-sectional means derived from the axisymmetric transport equation are contrasted with the reduced description to quantify the effect of transverse redistribution. The corresponding figures specify additional operating conditions, and numerical methods are summarised in Appendix~\ref{app:numerics}.

\subsection{Moment-based characterisation of pulse compression}
\label{sec:pulse_moments}

Following the moment-based framework of \citet{ChangSantiago2023}, we characterise pulse displacement and spreading via the axial moments of the cross-sectionally averaged concentration. The zeroth axial moment is defined as
\begin{equation}
M(T)=\int_0^1 C_0(Z,T)\,\mathrm{d}Z,
\end{equation}
which is proportional to the total solute content within the tube. The normalised first axial moment yields the axial centroid,
\begin{equation}
\mu(T)=\frac{1}{M(T)}\int_0^1 ZC_0(Z,T)\,\mathrm{d}Z.
\end{equation}
The axial variance quantifies the extent of solute spreading about the centroid:
\begin{equation}
\sigma_Z^2(T)=\frac{1}{M(T)}
\int_0^1 [Z-\mu(T)]^2 C_0(Z,T)\,\mathrm{d}Z.
\end{equation}
The corresponding standard deviation, $\sigma_Z(T)$, characterises the pulse width.

The ratio
\[
W(T)=\frac{\sigma_Z(T)}{\sigma_Z(0)},
\]
is the normalised pulse width: $W < 1$ indicates net contraction relative to the initial pulse, whereas $W > 1$ indicates net broadening. The sign of $\mathrm{d} \sigma_Z^2 / \mathrm{d}T$ determines instantaneous contraction or broadening and need not coincide with the sign of $W - 1$.

Moment-based diagnostics are evaluated over intervals in which more than 99.5\% of the initial solute mass remains within the domain, and the centroid is separated by more than four instantaneous standard deviations from either axial boundary. These criteria effectively limit boundary effects, which could otherwise obscure the distinction between pulse compression and solute loss through the axial boundaries.

Appendix~\ref{app:variance_balance} presents the finite-domain balances for the zeroth, first, and second axial moments, explicitly accounting for the axial-boundary contribution. Neglecting this contribution over intervals where axial-boundary effects are insignificant yields the approximation
\begin{equation}
\frac{1}{2}\frac{\mathrm d\sigma_Z^2}{\mathrm dT}
\simeq\underbrace{\left\langle(Z-\mu)\mathcal A_{\mathrm{eff}}\right\rangle_C}_{\text{advection}}
+\underbrace{\epsilon^2\left[\langle D_{\mathrm{eff}}\rangle_C+
\left\langle(Z-\mu)\frac{\mathrm{d}D_{\mathrm{eff}}}{\mathrm{d}Z}\right\rangle_C\right]}_{\text{dispersion}}.
\label{eq:results_variance_balance}
\end{equation}
Here, $\langle g \rangle_C = M^{-1} \int_0^1 g C_0\, \mathrm{d}Z$ denotes the concentration-weighted axial average of a function $g(Z, T)$, which is distinct from the cross-sectional average $\langle f \rangle$. The corresponding finite-domain identity, which accounts for the axial boundary contribution, is given by \eqref{eq:app_variance_with_boundaries}.


\section{Results and discussion}
\label{sec:physical_results}

\subsection{Hydraulic response to filtration and rheology}
\label{sec:hydraulic_response}
The hydraulic response couples the nonlinear pressure–flow relation~\eqref{eq:mean_velocity} to solvent loss through the wall, as described by equation~\eqref{eq:mean_velocity_gradient}. Together with the boundary conditions $P(0) = 1$ and $P(1) = 0$, these relations determine how filtration redistributes the axial flow and modifies the velocity field encountered by the solute pulse.

\begin{figure}[htbp]
\centering
\includegraphics[width=.98\linewidth]{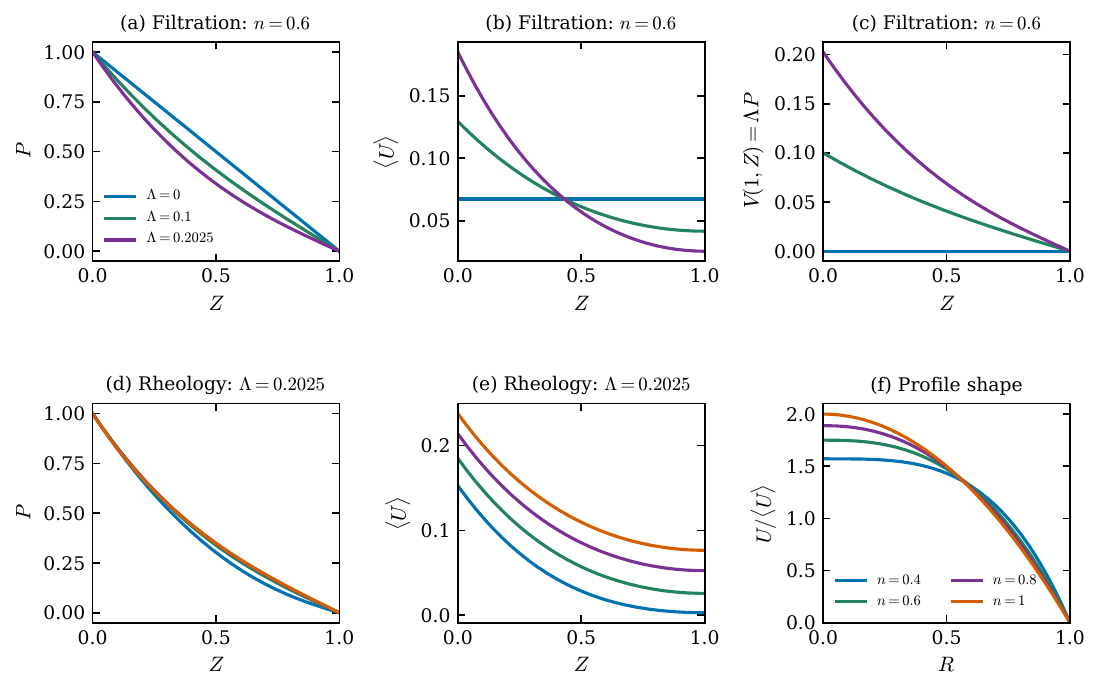}
\caption{Hydraulic response to wall filtration and power-law rheology. For fixed $n = 0.6$, panels (a–c) display the pressure $P(Z)$, the cross-sectional mean axial velocity $\langle U \rangle(Z)$, and the outward wall velocity $V(1, Z) = \Lambda P(Z)$ for $\Lambda = 0$, $0.1$, and $0.2025$. The legend in panel (a) applies to the entire top row. For fixed $\Lambda = 0.2025$, panels (d) and (e) compare the pressure and mean axial velocity for $n = 0.4$, $0.6$, $0.8$, and $1$. Panel (f) presents the corresponding normalised radial profiles, $U / \langle U \rangle = (3n + 1) (1 - R^{1 + 1 / n}) / (n + 1)$; its legend applies across the bottom row. The normalised shape is independent of $Z$ and $\Lambda$ within the lubrication solution. The mean velocity is proportional to the dimensionless axial volumetric flow rate. The pressure boundary conditions are $P(0) = 1$ and $P(1) = 0$ throughout. For $n < 1$, each displayed filtration strength is less than half of the corresponding critical value.}
\label{fig:hydraulic_response}
\end{figure}

We first examine the hydraulic fields that govern the effective transport coefficients. Figure~\ref{fig:hydraulic_response} distinguishes between variations in wall filtration at fixed rheology and changes in rheology at fixed dimensionless filtration strength. All cases depicted satisfy $\Lambda \leq 0.5 \Lambda_c(n)$ for $0 < n < 1$. The Newtonian case does not admit a finite $\Lambda_c$ in the present model and is included as a reference at the specified finite $\Lambda$.

At fixed pressure drop, increasing filtration alters both the inlet flow rate and its downstream decay rate, rather than simply removing a fixed fraction from an otherwise unchanged inlet flow. Panels (a–c) illustrate how outward solvent filtration modifies the axial pressure distribution and reduces the mean axial velocity downstream. The impermeable case exhibits linear pressure, uniform mean axial velocity, and zero transverse wall velocity. The rheological sweep in (d) and (e) modifies the nonlinear pressure–flow relation and, consequently, the axial flow distribution. Panel (f) isolates the effect of profile shape: decreasing $n$ flattens the radial velocity profile at a given local mean velocity. Thus, changes in axial flow and radial profile shape can be independently identified prior to assessing their combined effect on pulse deformation.

\subsection{Filtration-induced axial deceleration}
\label{sec:flow_physics}
In the reference pulse comparisons, $Pe = 14$, $\epsilon = 0.002$, $Z_0 = 0.25$, and $\sigma_0 = 0.06$ are held constant, while $n$ and $\Lambda$ are varied as indicated in each figure. The additional steady-coefficient sweep below varies $Pe$ at fixed hydraulic parameters. Dimensional reference scales are provided in Table~\ref{tab:dimensional_reference}. This protocol fixes the dimensionless transport parameters rather than the mean flow: the local mean-flow Péclet number is $Pe_m(Z) = Pe \langle U \rangle(Z)$, which varies with the hydraulic response. For a given geometry and $D$, a fixed $Pe$ determines $\mathcal{V}$, but the pressure $p_0 = m \ell \mathcal{V}^n / a^{n+1}$ and membrane conductivity $\mathcal{L}_w = \Lambda \epsilon \mathcal{V} / p_0$ depend on the constitutive properties. Thus, the parameter sweep does not hold the dimensional pressure, membrane conductivity, or inlet flow constant. The consistency index $m$ has dimension-dependent units, so numerical equality of $m$ across different indices does not constitute a material-equivalent comparison.

The axial gradient of the mean flow velocity is central to the mechanism of pulse compression. For $\Lambda=0$, $P=1-Z$, and $\langle U\rangle$ is constant. With outward filtration, however, continuity instead requires
\begin{equation}
\frac{\mathrm d\langle U\rangle}{\mathrm dZ}=-2\Lambda P<0.
\label{eq:flow_compression_mechanism}
\end{equation}
Consequently, fluid elements in the trailing portion of the pulse travel faster than those in the leading portion, resulting in axial contraction of the pulse, even in a tube of constant radius. This behaviour constitutes the kinematic analogue of the expansion-induced deceleration described by \citet{ChangSantiago2023}, but is here driven by wall filtration rather than by an increase in cross-sectional area. Solute retention is essential: while the solvent permeates the wall, the solute is retained and redistributed within the tube.

\begin{figure}[htbp]
\centering
\includegraphics[width=\linewidth]{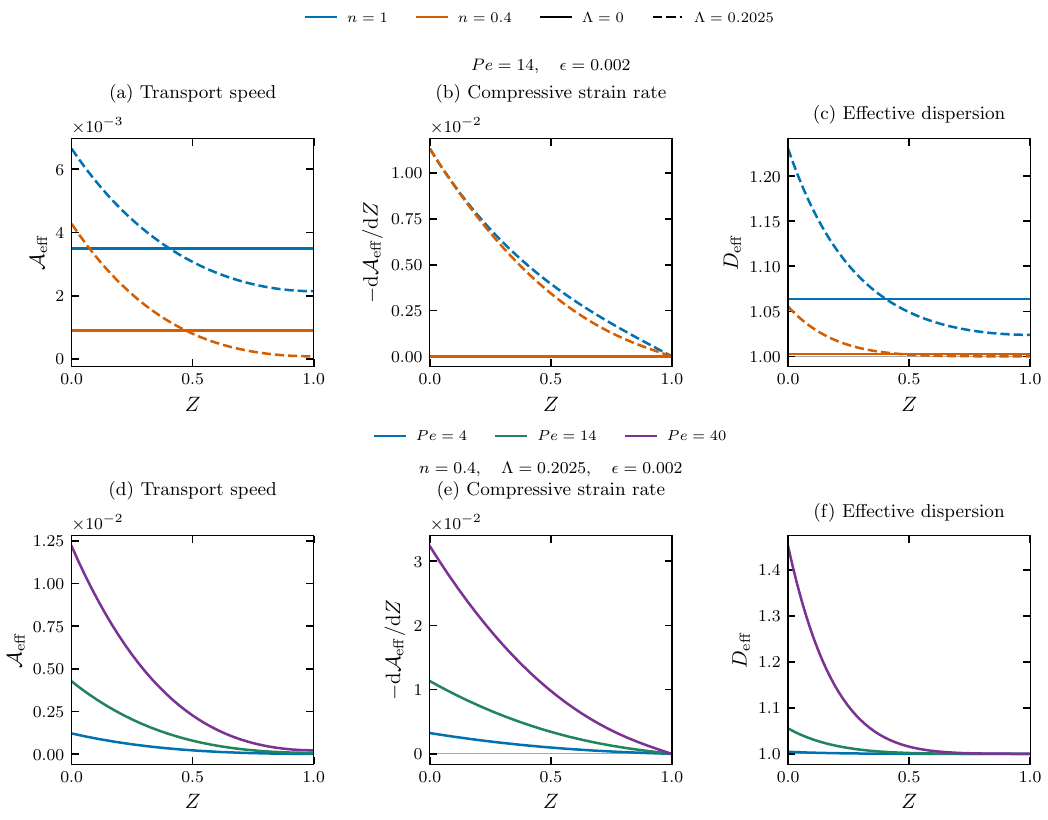}
\caption{Steady transport coefficients for $0 \leq Z \leq 1$, with $\epsilon = 0.002$ and pressure boundary conditions $P(0) = 1$, $P(1) = 0$. The columns display the effective axial velocity, the negative effective advection gradient (compressive strain rate), and the total effective dispersion. Panels (a–c) provide a reference comparison at $Pe = 14$ for all four combinations of $n = 1$, $0.4$ and $\Lambda = 0$, $0.2025$; colour distinguishes $n$ and line style distinguishes $\Lambda$. Panels (d–f) present a Péclet number sweep, $Pe = 4$, $14$, and $40$, at fixed $n = 0.4$ and $\Lambda = 0.2025$; the legend above (d–f) applies across the bottom row. The permeable shear-thinning cases satisfy $\Lambda / \Lambda_c(0.4) \simeq 0.455 < 0.5$. Positive $- \mathrm{d}\mathcal{A}_{\mathrm{eff}} / \mathrm{d}Z$ denotes deceleration, and both zero strain and $D_{\mathrm{eff}} = 1$ serve as reference levels. }
\label{fig:hydrodynamic_comparison}\label{fig:coefficient_comparison}
\end{figure}

Figure~\ref{fig:hydrodynamic_comparison}(a–c) presents the hydraulic response and associated transport coefficients experienced by the solute. In the absence of filtration, both the effective velocity and the dispersion coefficient are uniform, and the strain rate vanishes. Filtration increases the inlet flow under a prescribed pressure drop, but reduces the flow downstream. Consequently, the permeable and impermeable velocity profiles intersect, and solvent extraction does not produce a uniform reduction in transport along the tube. The Taylor contribution similarly decreases downstream as $\langle U \rangle^2$, such that an inlet value of $D_{\mathrm{eff}}$ does not characterise the spreading experienced over the entire trajectory.

Power-law rheology modifies both contributions. A reduction in $n$ leads to a flatter normalised radial velocity profile and reduces the factor $f(n) = n^2 / [2 (3n + 1)(5n + 1)]$; for a fixed local mean velocity, $f(0.4)/f(1) \simeq 0.582$. Across the operational regime considered, the mean flow also varies, producing a greater change in dispersion than that predicted by the profile factor alone. At $Z = 0.25$, the two permeable reference cases give $D_{\mathrm{eff}} \simeq 1.101$ for $n = 1$ and $1.013$ for $n = 0.4$, with comparable strain rates. Thus, rheology-dependent variations in pulse evolution should be interpreted as arising from the combined effect of compression and dispersion, rather than being attributed solely to modifications in the velocity-profile shape.

Panels (d–f) illustrate the effect of $Pe$ at fixed $n$, $\Lambda$, and prescribed pressure boundary conditions. Consequently, $P(Z)$ and $\langle U \rangle(Z)$ are independent of $Pe$. The leading mean-flow contributions to velocity and strain are proportional to $\epsilon Pe$, whereas the Taylor contribution, $D_{\mathrm{eff}} - 1$, and the wall correction scale as $Pe^2$. Increasing $Pe$ from 14 to 40 thus multiplies the Taylor contribution by $(40/14)^2 \simeq 8.16$, relative to a factor of $40/14 \simeq 2.86$ for the effective axial compression rate. The wall term marginally reduces the latter scaling. Increasing $Pe$ enhances both compressive advection and Taylor dispersion; thus, stronger advection does not necessarily yield greater pulse contraction. This interplay accounts for the intermediate-$Pe$ range favourable to contraction, as examined in Section~\ref{sec:contraction_threshold}.

The effective speed does not necessarily decrease along the entire tube. At $Z = 1$, the mean-flow gradient vanishes, and~\eqref{eq:kw_monotone} yields $\left.\mathrm{d} A_{\mathrm{eff}} / \mathrm{d} Z \right|_{Z = 1} > 0$ for $Pe > 0$ and $\Lambda > 0$. This results in a local increase in the effective transport speed near the outlet. The relative magnitude of this near-outlet increase grows with $Pe$, as the wall term is quadratic rather than linear in $Pe$. The compression relevant to the observed pulses occurs upstream of this outlet region.

\subsection{Displacement and deformation of the retained pulse}
\label{sec:pulse_evolution}
\begin{figure}[htbp]
\centering
\includegraphics[width=.94\linewidth]{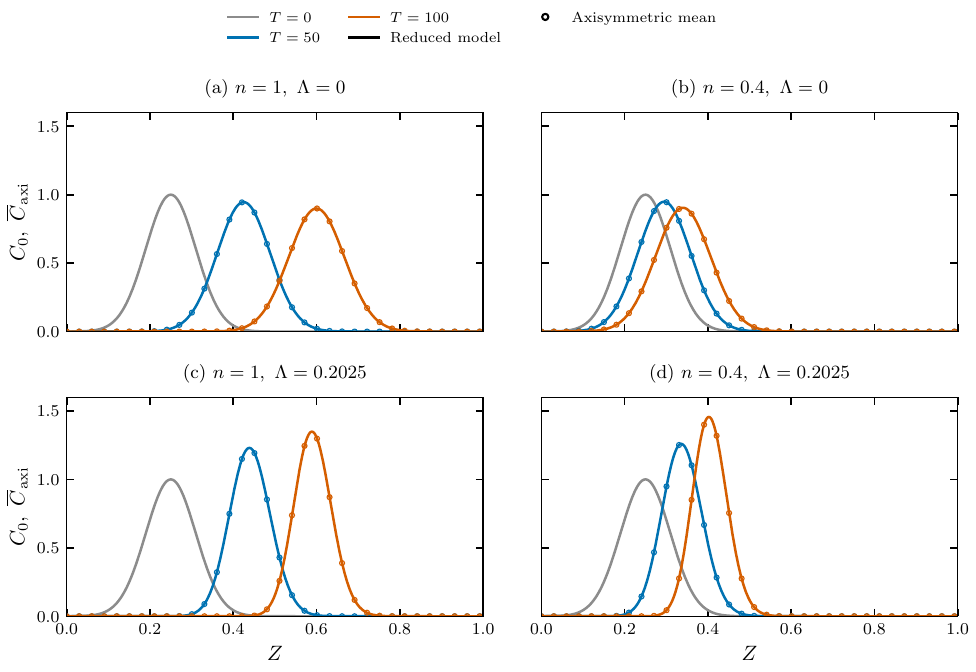}
\caption{Axial pulse evolution at $T = 0$, $50$, and $100$ for $Pe = 14$, $\epsilon = 0.002$, $Z_0 = 0.25$, and $\sigma_0 = 0.06$. Columns distinguish between Newtonian and shear-thinning fluids, while rows differentiate between impermeable and permeable walls. Solid lines represent the reduced-model profiles; open circles denote independently computed axisymmetric cross-sectional means. The profiles highlight downstream translation and spreading for impermeable walls, and axial compression under solvent filtration.}
\label{fig:axisymmetric_reference_nodal}\label{fig:concentration_comparison}
\end{figure}

Figure~\ref{fig:concentration_comparison} illustrates the modification of the solute distribution by kinematic competition. In the absence of filtration, uniform effective advection translates the pulse, while molecular and Taylor dispersion lead to its broadening. Under filtration, adjacent segments of the pulse are subject to different transport speeds. At $T = 50$ and $100$, the resulting pulse is narrower and exhibits a higher peak concentration than its impermeable counterpart. The increased peak arises from the redistribution of retained solute, rather than solute production or transwall passage. This narrowing is more pronounced for the shear-thinning case under the specified operating conditions.

The axisymmetric means accurately capture the pulse displacement and spreading. The following moment-based diagnostics quantify this evolution without presuming a Gaussian pulse shape.

\subsection{Variance identifies the balance between compression and spreading}
\label{sec:variance_physics}
\begin{figure}[htbp]
\centering
\includegraphics[width=.97\linewidth]{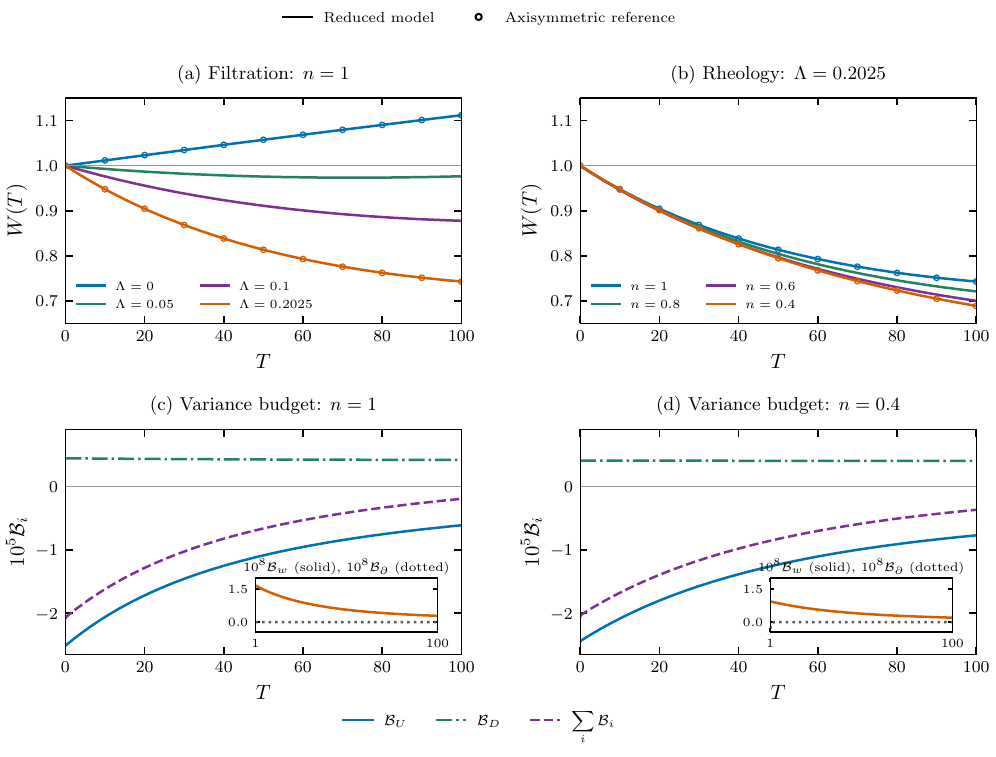}
\caption{Normalised pulse width and contributions to the axial variance growth rate for $Pe = 14$, $\epsilon = 0.002$, $Z_0 = 0.25$, and a common initial width $\sigma_0 = 0.06$. (a) Effect of dimensionless filtration strength at fixed $n = 1$, with $\Lambda = 0$, $0.05$, $0.1$, and $0.2025$. (b) Effect of rheology at a fixed filtration strength of $\Lambda = 0.2025$, with $n = 1$, $0.8$, $0.6$, and $0.4$. Solid lines indicate reduced-model histories of $W(T) = \sigma_Z(T) / \sigma_Z(0)$; open circles denote available axisymmetric reference moments for the endpoint cases. The $n = 1$, $\Lambda = 0.2025$ curve serves as the common reference between the two parameter sweeps. Panels (c) and (d) present the half-variance budget at $\Lambda = 0.2025$ for $n = 1$ and $0.4$, including mean-flow compression $\mathcal{B}_U$, dispersive spreading $\mathcal{B}_D$, and the full net rate, which includes $\mathcal{B}_w$ and $\mathcal{B}_{\partial}$. Insets display the minor wall correction and the axial-boundary contribution for $T \geq 1$. The level $W = 1$ marks the initial width, and zero in panels (c) and (d) separates instantaneous growth from contraction.}
\label{fig:width_variance_comparison}
\end{figure}

Figure~\ref{fig:width_variance_comparison} illustrates two parameter variations using a common initial pulse. Panel (a) demonstrates that increasing $\Lambda$ from the impermeable case progressively reduces $W(T)$ over the interval shown, highlighting the competition between filtration-induced deceleration and dispersive spreading. Panel (b) compares rheological effects at a fixed dimensionless filtration strength, $\Lambda = 0.2025$: decreasing $n$ results in greater cumulative narrowing within this parameter family. At $T = 100$, the endpoint width ratios are approximately $W = 0.743$ for $n = 1$ and $W = 0.689$ for $n = 0.4$. The common reference curve links the two parameter sweeps.

The variance identity~\eqref{eq:results_variance_balance} quantifies the interplay between flow-induced compression and dispersive spreading. The covariance $\langle (Z - \mu) \mathcal{A}_{\mathrm{eff}} \rangle_C$ becomes negative when faster solute from upstream converges upon slower solute downstream. For the pulses shown, the dispersive term—including the contribution from $\frac{\mathrm{d} D_{\mathrm{eff}}}{\mathrm{d} Z}$—opposes this compression. Panels (c) and (d) demonstrate that the negative mean-flow contribution outweighs the positive dispersive contribution for both rheologies. The narrower shear-thinning pulse arises from the cumulative effect of this balance. Notably, a pulse narrower than its initial width may already be broadening instantaneously: it is the sign of the net budget, rather than the absolute value of the width, that determines this distinction.

\subsubsection{The two effects of wall filtration}
\label{sec:advective_budget_decomposition}
The effective advection coefficient, as defined by equation~\eqref{eq:effective_advection}, comprises contributions from mean-flow transport and a wall-filtration correction. Their distinct roles in the exact half-variance budget for a finite domain are
\begin{equation}
 \frac{1}{2}\frac{\mathrm{d}\sigma_Z^2}{\mathrm{d}T}
 =\mathcal{B}_U+\mathcal{B}_w+\mathcal{B}_D+\mathcal{B}_{\partial},
 \label{eq:budget_decomposition}
\end{equation}
where the zeroth axial moment $M$, centroid $\mu$ and concentration-weighted average are defined in section~\ref{sec:pulse_moments}, and
\begin{align}\label{eq:budget_parts}
 \mathcal{B}_U&=\epsilon Pe\bigl\langle (Z-\mu)\langle U\rangle\bigr\rangle_C,\\
 \mathcal{B}_w&=-\epsilon^2\langle (Z-\mu)\mathcal K_w\rangle_C,\\
 \mathcal{B}_D&=\epsilon^2\left\langle D_{\mathrm{eff}}+(Z-\mu)\frac{\mathrm{d}D_{\mathrm{eff}}}{\mathrm{d}Z}\right\rangle_C.
\end{align}

The axial-boundary contribution $\mathcal{B}_{\partial}$ is retained as defined by the finite-domain identity in Appendix~\ref{app:variance_balance}. For clarity, it can be written as follows:
\begin{equation}
\begin{split}
\mathcal{B}_{\partial}={}&-\frac{1}{2M}
 \Bigl[\bigl((Z-\mu)^2-\sigma_Z^2\bigr)J\Bigr]_{0}^{1}\\
&-\frac{\epsilon^2}{M}\big[(Z-\mu)D_{\mathrm{eff}}C_0\big]_{0}^{1},
\end{split}
\label{eq:budget_boundary}
\end{equation}
where $J=\mathcal A_{\mathrm{eff}}C_0-\epsilon^2D_{\mathrm{eff}}\frac{\partial C_0}{\partial Z}$ denotes the total axial solute flux.

The hydrodynamic continuity relation, $\frac{\mathrm{d}\langle U \rangle}{\mathrm{d}Z} = -2 \Lambda P$, demonstrates that the mean axial velocity decreases monotonically for outward filtration ($\Lambda > 0$) and positive internal pressure. Consequently, $\mathcal{B}_U \leq 0$ for any non-negative pulse: the mean-flow contribution is intrinsically compressive. In contrast, combining the explicit expressions for $\mathcal{K}_w$ and $\langle U \rangle$ yields
\begin{equation}
 \mathcal K_w=\frac{Pe^2 n\Lambda}{2(5n+1)}P\langle U\rangle,
 \qquad
 \frac{\mathrm{d}\mathcal K_w}{\mathrm{d}Z}=\frac{Pe^2n\Lambda}{2(5n+1)}
 \left(\frac{\mathrm{d}P}{\mathrm{d}Z}\langle U\rangle-2\Lambda P^2\right)<0.
 \label{eq:kw_monotone}
\end{equation}
In the pressure-driven positive-throughflow regime, $\frac{\mathrm{d}P}{\mathrm{d}Z} < 0$ and $\langle U \rangle > 0$. Consequently, $\mathcal{B}_w \geq 0$: the filtration-induced correction to the effective axial advection opposes, rather than reinforces, the compression generated by the downstream decrease in mean flow. This contribution originates from the transverse redistribution of retained solute caused by solvent filtration, rather than from solute transport across the wall.

Although the term $-\epsilon^2 \mathcal{K}_w$ reduces the effective advection coefficient, its decreasing magnitude downstream contributes positively to the variance growth rate and thus counteracts compression. The insets in figure~\ref{fig:width_variance_comparison} indicate that this contribution is of order $10^{-8}$ in the reference cases, whereas the dominant mean-flow contribution is of order $10^{-5}$. Consequently, mean-flow deceleration dominates over the wall-induced correction in these reference scenarios. The complete budget also retains the axial-boundary contribution.

In Appendix~\ref{app:profile_criterion}, the interior balance is reformulated as $1 - \beta Pe + \alpha Pe^2$ for a prescribed concentration profile. This formulation separates molecular spreading and compression from the quadratic Taylor and wall contributions, yielding $\Theta > 1$ and the interval $Pe_- < Pe < Pe_+$ when $\alpha > 0$ and axial-boundary effects are negligible. This approach complements the evolving-width criterion by isolating the rheological dependence at a fixed profile; its supplementary parameter set is distinct from those of the reference histories.

\subsection{The evolving physical meaning of a local critical width}
\label{sec:contraction_threshold}
For an approximately symmetric pulse, whose width is small relative to the axial length scale across which the transport coefficients vary, the variance balance simplifies at leading order to
\begin{equation}
\frac12\frac{\mathrm d\sigma_Z^2}{\mathrm dT}
\simeq\left.\frac{\mathrm{d}\mathcal A_{\mathrm{eff}}}{\mathrm{d}Z}\right|_{Z=\mu}\sigma_Z^2
+\epsilon^2D_{\mathrm{eff}}(\mu).
\label{eq:local_variance_threshold_balance}
\end{equation}
The first term is proportional to the pulse variance: a broader pulse spans a greater velocity difference and experiences stronger compression. The second term represents the leading positive dispersive contribution. Higher-order derivatives of the coefficients and higher moments of the pulse are neglected at this order. When $\left.\frac{\mathrm{d}\mathcal{A}_{\mathrm{eff}}}{\mathrm{d}Z}\right|_{Z=\mu}<0$, the balance yields
\begin{equation}
\sigma_{Z,\mathrm{crit}}^2\simeq
\frac{\epsilon^2D_{\mathrm{eff}}(\mu)}{-\left.\frac{\mathrm{d}\mathcal A_{\mathrm{eff}}}{\mathrm{d}Z}\right|_{Z=\mu}}.
\label{eq:critical_pulse_width}
\end{equation}
Within the local approximation, pulses wider than the critical width undergo instantaneous contraction, whereas narrower pulses undergo spreading. This is an evolving balance that is evaluated at the centroid, rather than being a time-independent classification of the initial pulse. This distinction is similar to the transient negative-variance-growth regimes of \citet{ChangSantiago2023}, although the geometric slope is replaced here by the filtration-induced gradient of effective transport speed.

\begin{figure}[htbp]
\centering
\includegraphics[width=.83\linewidth]{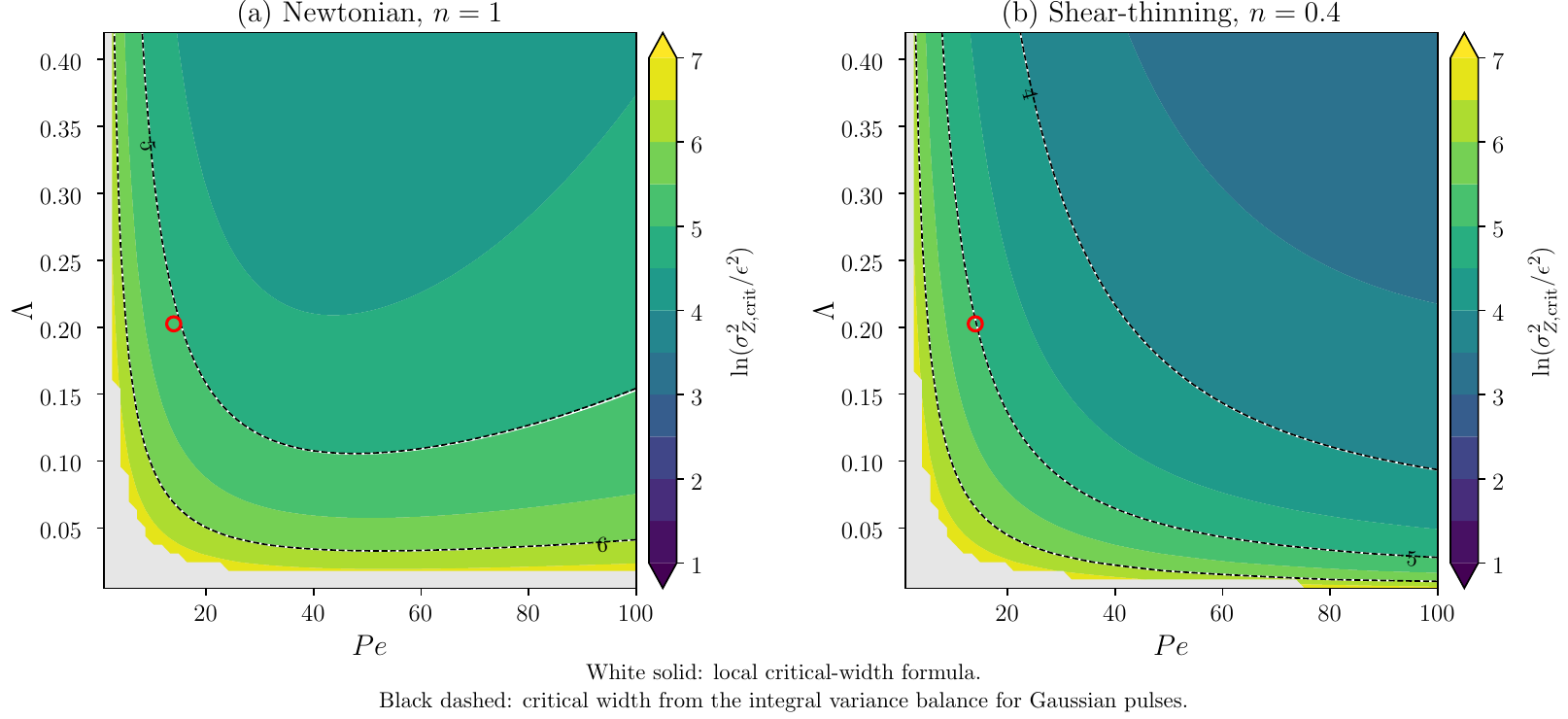}\par\vspace{3mm}
\includegraphics[width=.88\linewidth]{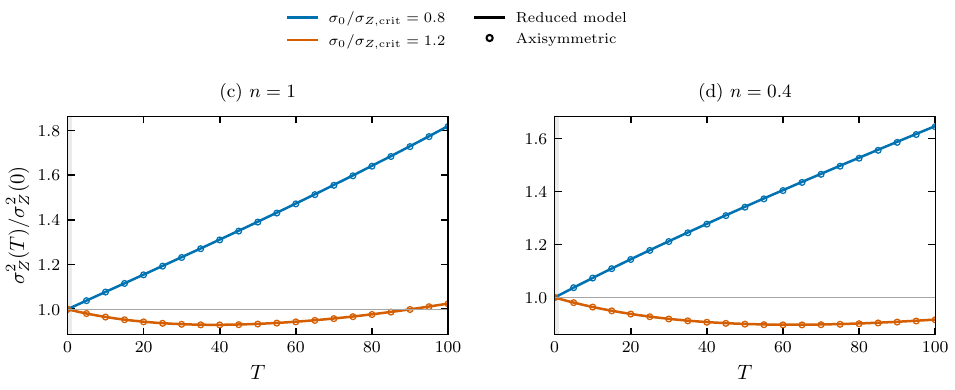}
\caption{Conditions for instantaneous contraction. (a, b) Maps of $\ln(\sigma_{Z,\mathrm{crit}}^2 / \epsilon^2)$ at $\epsilon = 0.002$ and $Z_0 = 0.25$ for $n = 1$ and $n = 0.4$. White solid contours correspond to the local formula~\eqref{eq:critical_pulse_width}; black dashed contours are derived from the integral variance balance for prescribed Gaussian profiles. Grey regions indicate pulse widths that do not satisfy the specified boundary-distance criterion. Red circles denote $Pe = 14$, $\Lambda = 0.2025$. (c, d) Normalised variance histories at this operating point, initiated from $\sigma_0 / \sigma_{Z,\mathrm{crit}} = 0.8$ and $1.2$, with each rheology's integral threshold. Solid lines represent reduced-model histories, and open circles denote axisymmetric moments. Shaded regions denote $0 \leq T < 1$, a selected initial adjustment window. The histories extend to $T = 100$. Narrower pulses exhibit broadening, whereas wider pulses initially contract and subsequently broaden; the reduced-model minima are attained near $T = 38.7$ for $n = 1$ and $T = 62.5$ for $n = 0.4$. }
\label{fig:contraction_threshold}
\end{figure}

Figure~\ref{fig:contraction_threshold}(a, b) demonstrates that the critical width does not necessarily vary monotonically with $Pe$. Stronger advection increases compressive deceleration, it simultaneously amplifies the Taylor contribution. This competition yields an intermediate $Pe$ range that is favourable to contraction for a prescribed pulse width. The maps encompass $1 \leq Pe \leq 100$ and $0.005 \leq \Lambda \leq 0.42$, remaining below $\Lambda_c(0.4) \simeq 0.446$; grey regions indicate thresholds excluded by the constraint $4 \sigma_Z \leq \min(Z_0, 1 - Z_0)$, rather than confirming the absence of contraction for all possible profiles. Notably, the hydrodynamic throughflow limit and the transport critical width represent distinct restrictions.

For $Pe = 14$ and $\Lambda = 0.2025$, the critical pulse widths determined from the integral variance balance for Gaussian concentration profiles are approximately $0.02521$ for $n = 1$ and $0.02452$ for $n = 0.4$. Panels (c) and (d) illustrate the corresponding evolution of axial variance, as predicted by the reduced model and axisymmetric simulations. For both rheologies, pulses initially wider than their respective critical widths contract following an initial adjustment, whereas narrower pulses exhibit spreading. Over the extended interval $0 \leq T \leq 100$, the initially contracting pulses attain a minimum variance before subsequently spreading. As the initial widths differ between the two rheologies, these panels serve to evaluate the contraction criterion, rather than directly compare rheological effects at identical initial widths. The close agreement of the two critical widths further indicates that the greater cumulative narrowing observed for the $\sigma_0 = 0.06$ reference pulse does not necessarily correspond to a proportionally large change in the threshold for instantaneous contraction.

As the pulse moves and its width evolves, both terms in~\eqref{eq:local_variance_threshold_balance} vary accordingly. Consequently, a pulse may transition from contraction to spreading as its centroid and width change over time. The threshold maps must therefore be interpreted with respect to a specified pulse position and concentration profile. The Gaussian profiles define the initial contraction threshold; subsequent deformation alters the balance as the pulse progresses downstream.

\subsection{Observed contraction intensity, reversal and solvent throughflow}
\label{sec:contraction_performance}
The instantaneous threshold does not specify the extent of narrowing achieved or the time at which contraction terminates. Both performance panels are restricted to $\Lambda \leq 0.5 \Lambda_c(n)$ for $0 < n < 1$, employing the rheology-dependent critical value from Appendix~\ref{app:critical_filtration}. For $n = 1$, no finite critical value exists, and the displayed range remains $0 \leq \Lambda \leq 0.42$. Accordingly, we track a common initial pulse with $\sigma_0 = 0.06$, $Z_0 = 0.25$, $Pe = 14$, and $\epsilon = 0.002$ over a two-parameter family of rheology and filtration strengths. We define the minimum observed width as
\begin{equation}
 W_{\min,\mathrm{obs}}=\min_{0\leq T\leq100}W(T),
 \label{eq:observed_minimum_width}
\end{equation}
restricted to trajectories that satisfy the solute-retention and boundary-distance criteria. We report a contraction-to-spreading transition time, $T_{\mathrm{rev}}$, only when an interior minimum is observed after the selected initial adjustment interval ($T < 1$). A trajectory whose minimum occurs at the final observation time is classified as exhibiting no observed reversal, rather than being assigned $T_{\mathrm{rev}} = 100$.

\begin{figure}[htbp]
\centering
\includegraphics[width=.97\linewidth]{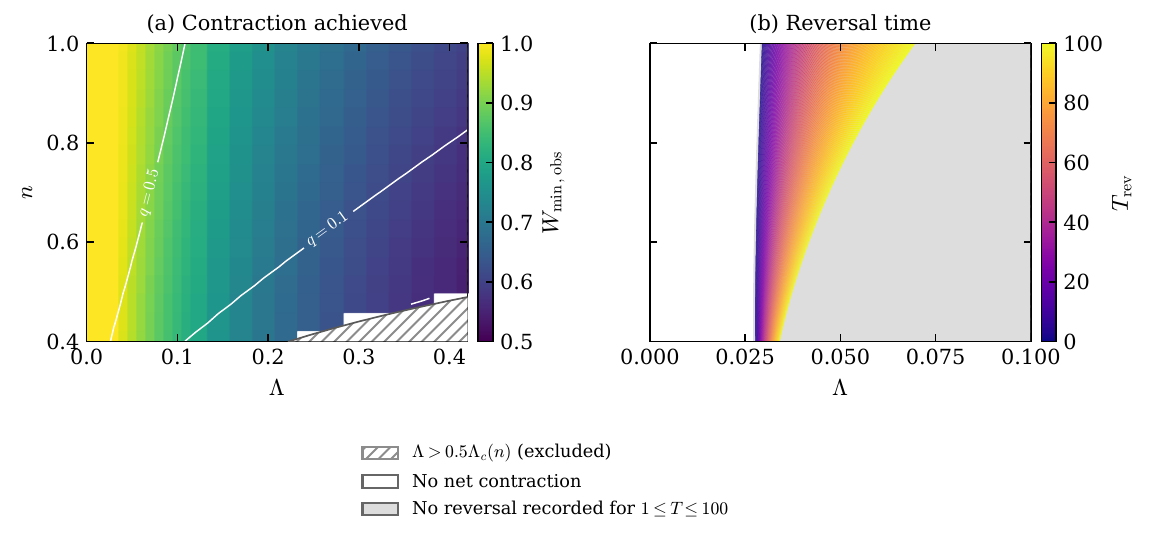}
\caption{Minimum observed normalised pulse width and reversal for a common initial width, $\sigma_0 = 0.06$, at $Pe = 14$, $\epsilon = 0.002$, and $Z_0 = 0.25$. (a) Minimum normalised width, $W_{\min,\mathrm{obs}}$, over $0 \leq T \leq 100$. White contours indicate the solvent-throughflow fraction, $q = Q_{\mathrm{out}} / Q_{\mathrm{in}}$, as evaluated from the steady hydraulic solution. (b) Reversal time, $T_{\mathrm{rev}}$, for trajectories exhibiting an interior minimum after $T = 1$, magnifying the weak-filtration region $0 \leq \Lambda \leq 0.1$. Grey denotes contraction without a recorded reversal in $1 \leq T \leq 100$; white denotes no net contraction below the initial width. Both panels impose $\Lambda \leq 0.5 \Lambda_c(n)$ for $n < 1$. Hatching identifies the excluded region. Positive outlet throughflow is preserved throughout; the maps characterise the reduced-model response before axial-boundary effects become significant.}
\label{fig:contraction_performance}
\end{figure}

Figure~\ref{fig:contraction_performance}(a) quantifies accumulated deformation, rather than the initial sign of the variance growth rate. The throughflow contours indicate how the attainable pulse narrowing corresponds to the residual solvent flow at the outlet. This ratio is a hydraulic quantity and does not represent the fraction of solute that is retained or delivered. For example, for $n = 1$, $\Lambda = 0.05$ yields $W_{\min,\mathrm{obs}} \simeq 0.9733$ and $T_{\mathrm{rev}} \simeq 72.8$. For $n = 0.4$, $\Lambda = 0.22$—which is below $0.5 \Lambda_c(0.4) \simeq 0.22275$—the observed width attains $0.6712$ at $T = 100$, with $q \simeq 0.01350$ and no observed reversal. This example demonstrates that pronounced narrowing alone does not guarantee favourable solvent delivery through the outlet. Panel (b) magnifies the weak-filtration region and differentiates between a contraction stage followed by broadening and contraction that persists throughout the observation window. For the stronger filtration cases, contraction persists throughout $0 \leq T \leq 100$, implying that the observation window captures only the compressive stage. Unlike the threshold-bracketing trajectories of figure~\ref{fig:contraction_threshold}, this comparison uses the same initial width for each rheology.

\FloatBarrier
\section{Conclusions}

Solute pulse compression within a semipermeable circular tube results from the interplay between filtration-induced axial flow deceleration and dispersive spreading. Employing lubrication theory and multiple-scale homogenisation, we have derived a one-dimensional macrotransport equation describing the transport of a retained solute in a power-law fluid. The axial variance balance distinguishes the compressive contribution of mean-flow deceleration from dispersive spreading and wall-induced concentration redistribution. Under the reference conditions considered, mean-flow deceleration dominates pulse compression, while the redistribution correction acts in opposition.

Power-law rheology modifies both the hydraulic response and the effective transport coefficients. At fixed local mean velocity, shear thinning reduces the velocity-profile contribution to Taylor dispersion. However, under the prescribed dimensionless operating conditions, rheology also influences the axial velocity field, and pulse evolution reflects the combined effects of compression and dispersion. At $T = 100$, the normalised pulse width is approximately $0.689$ for $n = 0.4$, compared to $0.743$ for the Newtonian case. These findings indicate greater cumulative narrowing for the shear-thinning reference case, without establishing a universal dependence on the flow behaviour index.

Within the local approximation, the variance balance yields a critical pulse width delineating instantaneous contraction from spreading. Pulses exceeding this threshold contract, whereas narrower pulses exhibit spreading. As the pulse position and width evolve, an initially contracting pulse may attain a minimum width before subsequently spreading. Comparisons with axisymmetric convection–diffusion simulations corroborate the predicted contraction thresholds. An increase in the Péclet number augments both compressive advection and Taylor dispersion, resulting in a non-monotonic dependence of the critical width on $Pe$.

Filtration imposes an independent hydraulic constraint. For shear-thinning fluids, positive outlet throughflow requires $\Lambda < \Lambda_c(n)$, whereas the present model admits no corresponding finite limit for $n \geq 1$. This hydraulic limit is distinct from the threshold for instantaneous pulse contraction. Substantial pulse narrowing may occur even when the outlet solvent flow rate constitutes only a small fraction of the inlet value. Accordingly, pulse narrowing should be evaluated alongside the remaining solvent throughflow and the duration of the contraction stage.

\section*{Acknowledgements}
\label{sec:acknowledgements}
G. Covarrubias acknowledges a Ph.D. fellowship from the Secretaría de Ciencia, Humanidades, Tecnología e Innovación (SECIHTI) and support from the Instituto Politécnico Nacional through its BEIFI programme. O. Bautista and J. Arcos acknowledge financial support from the Secretaría de Investigación y Posgrado of the Instituto Politécnico Nacional, Mexico, through research grants 20260368 and 20260435.



\begin{appen}
\renewcommand{\theHsection}{appendix.\Alph{section}}
\renewcommand{\theHequation}{\theHsection.\arabic{equation}}

\FloatBarrier
\section{Critical filtration parameter}\label{AppendixA}\label{app:critical_filtration}

The critical filtration parameter $\Lambda_c$ corresponds to the limiting state of vanishing mean axial velocity at the tube outlet. Since
$\langle U\rangle = n(3n+1)^{-1}\left(-\frac12\frac{\mathrm{d}P}{\mathrm{d}Z}\right)^{1/n}$,
the critical pressure profile satisfies
\begin{equation}
-\frac{n}{2(3n+1)}
\frac{\mathrm{d}}{\mathrm{d}Z}
\left[\left(-\frac12\frac{\mathrm{d}P}{\mathrm{d}Z}\right)^{1/n}\right]
=\Lambda_c P,
\qquad
P(0)=1,\quad P(1)=\left.\frac{\mathrm{d}P}{\mathrm{d}Z}\right|_{Z=1}=0,
\label{eq:critical_pressure}
\end{equation}
The outlet condition $\left.\frac{\mathrm{d}P}{\mathrm{d}Z}\right|_{Z=1}=0$ determines $\Lambda_c$;
it is not an additional boundary condition for arbitrary
$\Lambda$. We seek a profile with $P>0$ and $\frac{\mathrm{d}P}{\mathrm{d}Z}<0$ for
$0\leq Z<1$.

Multiplying \eqref{eq:critical_pressure} by $\mathrm{d}P/\mathrm{d}Z$ and integrating with respect to $Z$, subject to the outlet conditions, yields
\begin{equation}
\left(-\frac12\frac{\mathrm{d}P}{\mathrm{d}Z}\right)^{(n+1)/n}
=
\frac{(n+1)(3n+1)}{2n}\Lambda_c P^2.
\label{eq:critical_first_integral}
\end{equation}
Separating variables and integrating over the tube yields
\begin{equation}
\int_0^1 P^{-2n/(n+1)}\,\mathrm{d}P
=
2\left[
\frac{(n+1)(3n+1)}{2n}\Lambda_c
\right]^{n/(n+1)}.
\label{eq:critical_integral}
\end{equation}
For $0<n<1$, the integral evaluates to $(n+1)/(1-n)$, yielding
\begin{equation}
\Lambda_c(n)=
\frac{2n}{(n+1)(3n+1)}
\left[\frac{n+1}{2(1-n)}\right]^{(n+1)/n},
\qquad 0<n<1.
\label{eq:critical_filtration_appendix}
\end{equation}
Positive outlet throughflow corresponds to
$0\leq\Lambda<\Lambda_c(n)$, while
$\Lambda=\Lambda_c(n)$ marks vanishing outlet flow.
For $n\geq1$, the integral in
\eqref{eq:critical_integral} diverges, and no finite critical
filtration parameter arises through this mechanism within
the present model.

\FloatBarrier
\section{Derivation of the pulse-variance balance}
\label{app:variance_balance}

Following the classical method of moments of \citet{aris1956dispersion}, reviewed by \citet{young1991shear}, we derive the finite-domain variance balance for the present spatially varying transport coefficients while retaining the axial-boundary contributions. The methodological basis is provided by these references; the coefficient-dependent terms and finite-domain boundary contributions below are obtained directly from the present transport equation.

We first consider the expanded, non-conservative form of the retained-order macrotransport equation~\eqref{eq:macro_transport_advective_form}:
\begin{equation}
\frac{\partial C}{\partial T}+A \frac{\partial C}{\partial Z}+\frac{\mathrm{d}A}{\mathrm{d}Z}C=\epsilon^2 D \frac{\partial^2 C}{\partial Z^2}+ \epsilon^2  \frac{\mathrm{d}D}{\mathrm{d}Z}\frac{\partial C}{\partial Z}.
\label{eq:app_variance_expanded}
\end{equation}
Here $C=C_0$, $A=\mathcal A_{\mathrm{eff}}$, and $D=D_{\mathrm{eff}}$ are local abbreviations. The coefficients depend on $Z$ but not on $T$. In particular, the terms $\frac{\mathrm{d}A}{\mathrm{d}Z}C$ and $ \epsilon^2 \frac{\mathrm{d}D}{\mathrm{d}Z}\frac{\partial C}{\partial Z}$ must be retained when the coefficients vary axially.

Using the zeroth axial moment, centroid and variance defined in section~\ref{sec:pulse_moments}, we write $Q_j=\int_0^1Z^jC\,\mathrm dZ$, where $Q_j$ is the unnormalised axial moment of order $j=0,1,2$. Thus $Q_0=M$, $Q_1=M\mu$ and $Q_2=M(\mu^2+\sigma_Z^2)$. The notation $[f]_0^1=f(1,T)-f(0,T)$ denotes the difference between outlet and inlet values.

For a smooth weight $\phi(Z)$, integration by parts of the two advective terms gives
\begin{equation}
-\int_0^1\phi A \frac{\partial C}{\partial Z}\,\mathrm dZ
-\int_0^1\phi \frac{\mathrm{d}A}{\mathrm{d}Z}C\,\mathrm dZ
=-[\phi AC]_0^1+\int_0^1\frac{\mathrm{d}\phi}{\mathrm{d}Z}AC\,\mathrm dZ.
\label{eq:app_weighted_advection}
\end{equation}
The terms involving $\phi(\mathrm{d}A/\mathrm{d}Z)C$ cancel exactly. Similarly, integrating the second-derivative term twice and the coefficient-gradient term once yields
\begin{equation}
\begin{aligned}
 \int_0^1\phi D \frac{\partial^2 C}{\partial Z^2}\,\mathrm dZ
+   \int_0^1\phi \frac{\mathrm{d}D}{\mathrm{d}Z}\frac{\partial C}{\partial Z}\,\mathrm dZ
&=   \left[\phi D\frac{\partial C}{\partial Z}\right]_0^1-   \left[\frac{\mathrm{d}\phi}{\mathrm{d}Z}DC\right]_0^1\\
&\quad+   \int_0^1\left(\frac{\mathrm{d}^2\phi}{\mathrm{d}Z^2}D+\frac{\mathrm{d}\phi}{\mathrm{d}Z}\frac{\mathrm{d}D}{\mathrm{d}Z}\right)C\,\mathrm dZ.
\end{aligned}
\label{eq:app_weighted_diffusion}
\end{equation}
These integrations retain the contributions arising from axial variations in both transport coefficients.

Setting $\phi=1$ yields the zeroth-moment balance
\begin{equation}
\frac{\mathrm{d}M}{\mathrm{d}T}=-[AC]_0^1+ \epsilon^2  \left[D\frac{\partial C}{\partial Z}\right]_0^1.
\label{eq:app_inventory_balance}
\end{equation}
Now setting $\phi=Z$ gives the first-moment balance
\begin{equation}
\begin{aligned}
\frac{\mathrm{d}Q_1}{\mathrm{d}T}
&=-[ZAC]_0^1+ \epsilon^2  \left[ZD\frac{\partial C}{\partial Z}\right]_0^1-\epsilon^2  [DC]_0^1\\
&\quad+M\left(\langle A\rangle_C+ \epsilon^2 \left\langle \frac{\mathrm{d}D}{\mathrm{d}Z}\right\rangle_C\right).
\end{aligned}
\label{eq:app_first_moment}
\end{equation}
Finally, setting $\phi=Z^2$ yields the second-moment balance
\begin{equation}
\begin{aligned}
\frac{\mathrm{d}Q_2}{\mathrm{d}T}
&=-[Z^2AC]_0^1+\epsilon^2  \left[Z^2D\frac{\partial C}{\partial Z}\right]_0^1-2 \epsilon^2  [ZDC]_0^1\\
&\quad+2M\left(\langle ZA\rangle_C
+ \epsilon^2  \langle D\rangle_C+ \epsilon^2  \left\langle Z\frac{\mathrm{d}D}{\mathrm{d}Z}\right\rangle_C\right).
\end{aligned}
\label{eq:app_second_moment}
\end{equation}
Here $\langle g\rangle_C=M^{-1}\int_0^1gC\,\mathrm dZ$ denotes the concentration-weighted axial average. The contribution involving $\frac{\mathrm{d}D}{\mathrm{d}Z}$ therefore arises from integration by parts of the spatially varying dispersion terms, without introducing an axial-flux variable.

Differentiating the normalised variance, without assuming that \(M\) is constant
\begin{equation}
\frac12\frac{\mathrm d\sigma_Z^2}{\mathrm dT}
=\frac{\frac{\mathrm{d}Q_2}{\mathrm{d}T}-2\mu\frac{\mathrm{d}Q_1}{\mathrm{d}T}
+(\mu^2-\sigma_Z^2)\frac{\mathrm{d}M}{\mathrm{d}T}}{2M}.
\label{eq:app_normalised_variance}
\end{equation}
Substitution of~\eqref{eq:app_inventory_balance}--\eqref{eq:app_second_moment} gives
\begin{equation}
\begin{aligned}
\frac12\frac{\mathrm d\sigma_Z^2}{\mathrm dT}
&=\underbrace{\langle(Z-\mu)A\rangle_C}_{\text{Advective contribution}}\\
&\quad+\underbrace{\epsilon^2  \left(\langle D\rangle_C+\left\langle(Z-\mu)\frac{\mathrm{d}D}{\mathrm{d}Z}\right\rangle_C\right)}_{\text{Dispersive contribution}}
+\underbrace{\mathcal B_{\partial}}_{\text{Boundary contribution}},\\
\mathcal B_{\partial}
&=-\frac{1}{2M}\Bigl[\bigl((Z-\mu)^2-\sigma_Z^2\bigr)AC\Bigr]_0^1\\
&\quad+\frac{\epsilon^2  }{2M}\left[\bigl((Z-\mu)^2-\sigma_Z^2\bigr)D\frac{\partial C}{\partial Z}\right]_0^1
-\frac{\epsilon^2 }{M}\bigl[(Z-\mu)DC\bigr]_0^1.
\end{aligned}
\label{eq:app_variance_with_boundaries}
\end{equation}
Here, $\mathcal B_{\partial}$ represents the total contribution of the axial boundaries to one half of the variance growth rate. Its three terms account for advective transport, diffusive transport associated with the axial concentration gradient, and the additional boundary contribution involving $D_{\mathrm{eff}}(Z)C_0(Z,T)$. The normalisation of the variance also accounts for changes in the total amount of solute within the tube.

If the pulse remains sufficiently far from the ends that the relevant traces of $AC$, $D\frac{\partial C}{\partial Z}$ and $DC$ are negligible, then $\mathcal B_{\partial}$ is negligible and~\eqref{eq:app_inventory_balance} gives approximately constant $M$. The resulting approximation is
\begin{equation}
\frac12\frac{\mathrm d\sigma_Z^2}{\mathrm dT}
\simeq\left\langle(Z-\mu)\mathcal A_{\mathrm{eff}}\right\rangle_C
+\epsilon^2\left[\langle D_{\mathrm{eff}}\rangle_C
+\left\langle(Z-\mu)\frac{\mathrm{d}D_{\mathrm{eff}}}{\mathrm{d}Z}\right\rangle_C\right],
\label{eq:app_variance_negligible_boundaries}
\end{equation}
which is equation~\eqref{eq:results_variance_balance}. The condition $AC-\epsilon^2  D\frac{\partial C}{\partial Z}=0$ at an axial boundary alone does not eliminate the $DC$ boundary term. The solute-retention and boundary-distance criteria in section~\ref{sec:pulse_moments} are practical checks for interpreting the boundary-negligible balance, rather than rigorous estimates of $\mathcal B_{\partial}$.

\FloatBarrier
\section{Numerical implementation and scope of the comparisons}
\label{app:numerics}

The reduced and axisymmetric transport models employ a common steady lubrication velocity field, initial solute distribution, and axial boundary conditions. The hydraulic boundary-value problem is solved using adaptive collocation,
and analytical velocity profiles are used to determine the
transport coefficients. The numerical discretisations used for the one-dimensional macrotransport equation and the two-dimensional axisymmetric convection–diffusion equation are described below.

\subsection{Macrotransport integration}
\label{sec:numerical_macrotransport}

The one-dimensional macrotransport equation
\eqref{eq:macro_transport_advective_form} is solved numerically
subject to the following boundary conditions:
\begin{equation}
C_0(0,T)=0,
\qquad
\left.\frac{\partial C_0}{\partial Z}\right|_{Z=1}=0,
\label{eq:appC_macro_bc}
\end{equation}
and the initial condition
\begin{equation}
C_0(Z,0)=
\exp\left[-\frac{(Z-Z_0)^2}{2\sigma_0^2}\right].
\label{eq:appC_macro_ic}
\end{equation}

The governing equation is integrated using the Crank–Nicolson
scheme, which is second-order accurate in time, combined with
second-order centred finite differences in the spatial interior
\citep{anderson1995computational,chapra2010numerical}.

As the hydrodynamic field is steady, the spatially varying
effective transport coefficients are precomputed prior to time
integration and stored at the nodes of the discretised axial domain.
The pressure gradient and mean velocity are obtained directly from the
hydrodynamic solution. Spatial derivatives of the transport
coefficients are calculated using centred differences in the
interior and one-sided differences at the boundaries.

Expanding the derivatives in the one-dimensional macrotransport
equation yields
\begin{equation}
\begin{aligned}
\frac{\partial C_0}{\partial T}
&+
\mathcal A_{\mathrm{eff}}
\frac{\partial C_0}{\partial Z}
+
\frac{\mathrm d\mathcal A_{\mathrm{eff}}}{\mathrm dZ}C_0
\\
&=
\epsilon^2D_{\mathrm{eff}}
\frac{\partial^2C_0}{\partial Z^2}
+
\epsilon^2
\frac{\mathrm dD_{\mathrm{eff}}}{\mathrm dZ}
\frac{\partial C_0}{\partial Z}.
\end{aligned}
\label{eq:appC_macro_expanded}
\end{equation}

We introduce uniform spatial and temporal increments, $\Delta Z$
and $\Delta T$, respectively. The Crank–Nicolson discretisation
at axial node $i$ may be expressed as
\begin{equation}
\begin{aligned}
\frac{C_{0,i}^{m+1}-C_{0,i}^{m}}{\Delta T}
&+
\frac{\mathcal A_{\mathrm{eff},i}}{2}
\left(
\delta_Z C_{0,i}^{m+1}
+
\delta_Z C_{0,i}^{m}
\right)
\\
&+
\frac{1}{2}
\left(
\frac{\mathrm d\mathcal A_{\mathrm{eff}}}{\mathrm dZ}
\right)_i
\left(
C_{0,i}^{m+1}+C_{0,i}^{m}
\right)
\\
&=
\frac{\epsilon^2D_{\mathrm{eff},i}}{2}
\left(
\delta_Z^2 C_{0,i}^{m+1}
+
\delta_Z^2 C_{0,i}^{m}
\right)
\\
&\quad+
\frac{\epsilon^2}{2}
\left(
\frac{\mathrm dD_{\mathrm{eff}}}{\mathrm dZ}
\right)_i
\left(
\delta_Z C_{0,i}^{m+1}
+
\delta_Z C_{0,i}^{m}
\right).
\end{aligned}
\label{eq:appC_CN}
\end{equation}

Here, superscripts $m$ and $m+1$ denote the current and subsequenttime levels, whereas subscript $i$ designates the axial grid node. The operators $\delta_Z$ and $\delta_Z^2$ denote the standard second-order centred finite-difference approximations:
\begin{equation}
\delta_Z C_{0,i}
=
\frac{C_{0,i+1}-C_{0,i-1}}{2\Delta Z},
\qquad
\delta_Z^2 C_{0,i}
=
\frac{C_{0,i+1}-2C_{0,i}+C_{0,i-1}}
{(\Delta Z)^2}.
\label{eq:appC_CN_operators}
\end{equation}

Application of equation~\eqref{eq:appC_CN} at the interior
grid points yields a tridiagonal linear algebraic system,
which is solved via the Thomas algorithm.

A zero-concentration condition is imposed at the inlet,
whereas a zero axial concentration gradient at the outlet
is enforced using a ghost node.

The numerical implementation employs 1,000 uniformly distributed
axial nodes and a dimensionless time increment of
$\Delta T = 10^{-3}$.

Because the nodal discretisation is not expressed directly in conservative cell-flux form, the zeroth axial moment, $M(T)$, is monitored throughout the numerical integration to identify the influence of axial boundary transport on the computed pulse evolution.

\subsection{Axisymmetric integration}
\label{sec:numerical_2D_transport}

The complete two-dimensional axisymmetric convection–diffusion
equation~\eqref{eq:dimensionless_convection_diffusion} is solved
numerically over the dimensionless domain
$(R, Z) \in [0, 1] \times [0, 1]$.

The radial boundary conditions are given by
\begin{equation}
\left.\frac{\partial C}{\partial R}\right|_{R=0}=0,
\qquad
\left.\frac{\partial C}{\partial R}\right|_{R=1}
=
\epsilon Pe\,\Lambda P(Z)C(1,Z,T).
\label{eq:appC_axis_radial_bc}
\end{equation}

The axial boundary conditions are
\begin{equation}
C(R,0,T)=0,
\qquad
\left.\frac{\partial C}{\partial Z}\right|_{Z=1}=0.
\label{eq:appC_axis_axial_bc}
\end{equation}

The initial concentration distribution is
\begin{equation}
C(R,Z,0)=
\exp\left[-\frac{(Z-Z_0)^2}{2\sigma_0^2}\right].
\label{eq:appC_axis_ic}
\end{equation}

The initial condition specifies a Gaussian concentration pulse
centred at $Z_0$, with standard deviation parameter $\sigma_0$,
which is initially uniform in the radial direction.

The first radial boundary condition imposes axisymmetry at
the centreline, whereas the second enforces complete solute
retention at the semipermeable wall under outward solvent filtration.
The axial boundary conditions prescribe a zero inlet concentration and a
vanishing axial concentration gradient at the outlet.

The axisymmetric transport equation is integrated numerically using the
Peaceman–Rachford alternating-direction implicit (ADI) method,
with centred spatial finite differences
\citep{peaceman1955numerical,press2007numerical}.

Each time step, $\Delta T$, is divided into two successive
half-steps of duration $\Delta T/2$.
The first half-step treats the radial transport operator
implicitly, and the axial operator explicitly.
The second half-step reverses these roles.

Let $C_{j,i}^{m}$ denote the numerical concentration at
radial node $R_j$, axial node $Z_i$, and time level $m$.
The intermediate concentration is denoted by
$C_{j,i}^{m+1/2}$.

The first ADI half-step, implicit in the radial coordinate,
is expressed as follows:
\begin{equation}
\begin{aligned}
\frac{
C_{j,i}^{m+1/2}-C_{j,i}^{m}
}{\Delta T/2}
&+
\epsilon Pe\,V_{j,i}
\delta_R C_{j,i}^{m+1/2}
\\
&-
\delta_R^2 C_{j,i}^{m+1/2}
-
\frac{1}{R_j}
\delta_R C_{j,i}^{m+1/2}
\\
&=
-\epsilon Pe\,U_{j,i}
\delta_Z C_{j,i}^{m}
+
\epsilon^2\delta_Z^2 C_{j,i}^{m}.
\end{aligned}
\label{eq:appC_ADI_radial}
\end{equation}

The second ADI half-step, implicit in the axial coordinate,
is given by the following expression:
\begin{equation}
\begin{aligned}
\frac{
C_{j,i}^{m+1}-C_{j,i}^{m+1/2}
}{\Delta T/2}
&+
\epsilon Pe\,U_{j,i}
\delta_Z C_{j,i}^{m+1}
-
\epsilon^2\delta_Z^2 C_{j,i}^{m+1}
\\
&=
-\epsilon Pe\,V_{j,i}
\delta_R C_{j,i}^{m+1/2}
\\
&\quad+
\delta_R^2 C_{j,i}^{m+1/2}
+
\frac{1}{R_j}
\delta_R C_{j,i}^{m+1/2}.
\end{aligned}
\label{eq:appC_ADI_axial}
\end{equation}

The spatial difference operators that appear in
equations~\eqref{eq:appC_ADI_radial} and
\eqref{eq:appC_ADI_axial} are defined as follows:
\begin{equation}
\delta_R C_{j,i}
=
\frac{C_{j+1,i}-C_{j-1,i}}{2\Delta R},
\qquad
\delta_R^2 C_{j,i}
=
\frac{C_{j+1,i}-2C_{j,i}+C_{j-1,i}}
{(\Delta R)^2},
\label{eq:appC_radial_operators}
\end{equation}
and
\begin{equation}
\delta_Z C_{j,i}
=
\frac{C_{j,i+1}-C_{j,i-1}}{2\Delta Z},
\qquad
\delta_Z^2 C_{j,i}
=
\frac{C_{j,i+1}-2C_{j,i}+C_{j,i-1}}
{(\Delta Z)^2}.
\label{eq:appC_axial_operators}
\end{equation}

At the centreline, the apparent singularity in the cylindrical diffusion operator is resolved by application of L'Hôpital's rule in conjunction with the symmetry condition. The resulting regular limit is approximated by
\begin{equation}
\left.
\frac{1}{R}
\frac{\partial}{\partial R}
\left(
R\frac{\partial C}{\partial R}
\right)
\right|_{R=0}
\simeq
4\frac{C_{1,i}-C_{0,i}}{(\Delta R)^2}.
\label{eq:appC_axis_limit}
\end{equation}

At the semipermeable wall, the radial concentration
gradient is determined from the zero-total-normal-solute-flux
condition~\eqref{eq:appC_axis_radial_bc}, rather than
from a zero-gradient approximation.

The steady velocity fields, $U(R, Z)$ and $V(R, Z)$,
are obtained from the lubrication solution and
evaluated at each grid point prior to time integration.

Each ADI half-step reduces to a sequence of uncoupled
tridiagonal linear systems along the corresponding
coordinate lines. These systems are solved via
the Thomas algorithm.

After obtaining the two-dimensional concentration field,
the cross-sectional mean concentration is evaluated as follows:
\begin{equation}
\langle C\rangle(Z,T)
=
2\int_0^1 C(R,Z,T)R\,\mathrm dR.
\label{eq:appC_axis_average}
\end{equation}

The cross-sectional integral is evaluated using
the composite trapezoidal rule.

The numerical implementation employs a uniform, structured
mesh with $N_R = N_Z = 500$ nodes and a dimensionless
time increment $\Delta T = 10^{-2}$.

Unlike the reduced macrotransport equation, which
employs post-transient effective transport coefficients,
the axisymmetric formulation explicitly resolves the
initial transverse concentration adjustment associated
with the semipermeable wall.

\subsection{Temporal sensitivity and mesh independence analysis}
\label{sec:numerical_verification}

\begin{figure}
\centering
\includegraphics[width=0.65\textwidth]{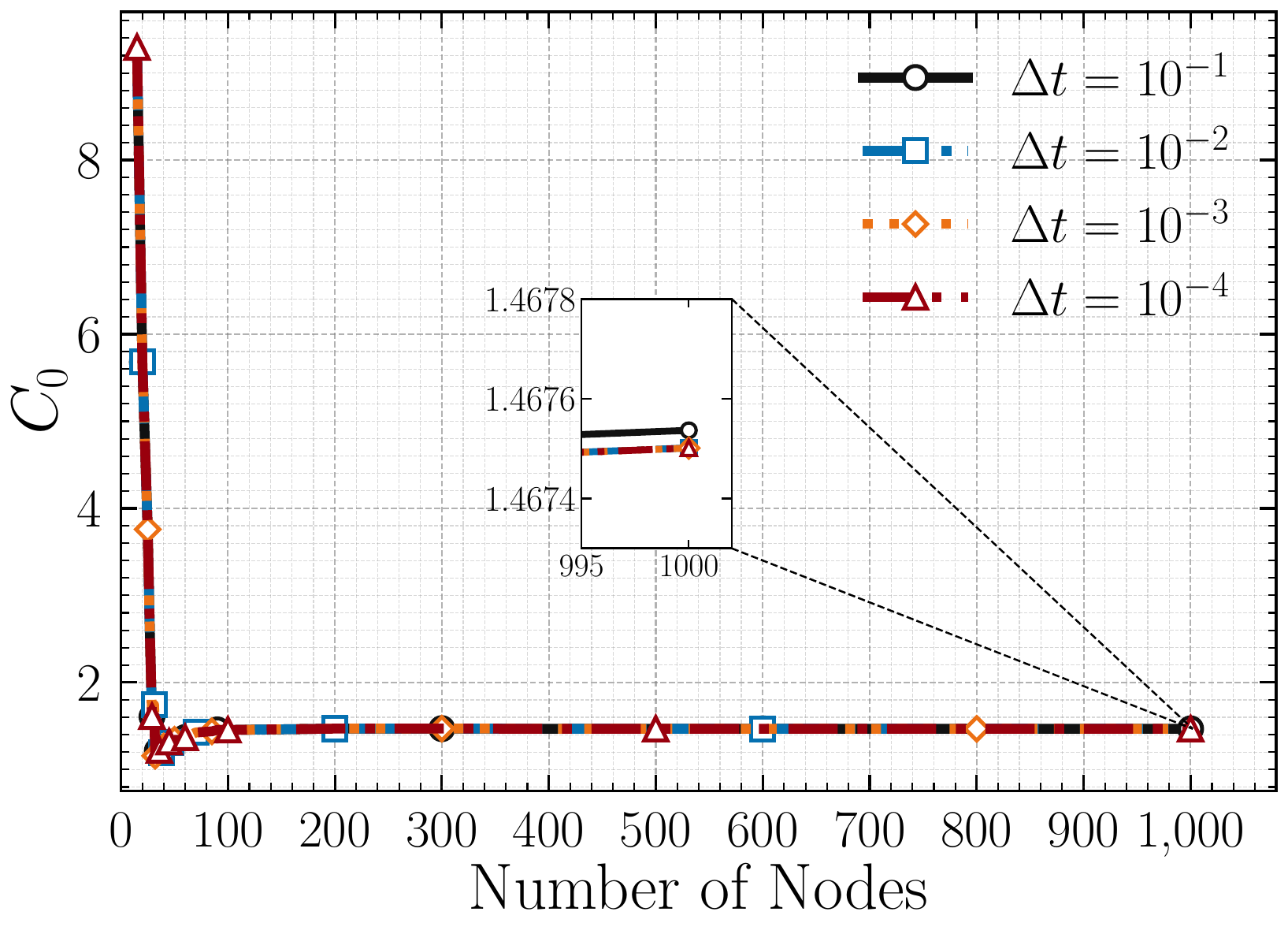}
\caption{The maximum mean concentration, $C_{0,\max}$, at $T = 100$, as a function of spatial grid resolution (number of nodes, $m$) for various temporal increments ($\Delta T = 10^{-1}, 10^{-2}, 10^{-3}$, and $10^{-4}$).}
\label{fig:mesh_convergence}
\end{figure}

Figure~\ref{fig:mesh_convergence} presents the grid convergence behaviour for the maximum mean concentration, $C_{0,\max}$, computed at dimensionless time $T = 100$. The physical and rheological parameters selected for this validation baseline correspond to a power-law fluid with $n = 1.0$, filtration parameter $\Lambda = 0.6$, Péclet number $Pe = 17$, aspect ratio $\epsilon = 0.0025$, and a Gaussian initial solute distribution centred at $Z_0 = 0.2$ with standard deviation $\sigma_0 = 0.04$. This configuration was selected to verify the spatial and temporal independence of the numerical solutions reported in this study.

The temporal sensitivity of the implemented Crank–Nicolson scheme is first examined via the curves shown in figure~\ref{fig:mesh_convergence}. For the coarsest temporal increment considered ($\Delta T = 10^{-1}$), the computed maximum mean concentration exhibits a slight upward offset across all nodal distributions, although it remains close to the refined solutions. By contrast, as the time increment is reduced, the curves corresponding to the remaining time-steps become virtually indistinguishable. Specifically, for $\Delta T \leq 10^{-2}$, the spatial profiles collapse onto a single curve, demonstrating that the numerical algorithm rapidly achieves temporal independence.

To rigorously quantify this behaviour, a convergence analysis was conducted by evaluating the relative error between consecutive temporal discretisations on the finest spatial mesh ($m = 1000$). The relative error is defined as $e_{\mathrm{rel}}^{ji} = |(C_{0,i} - C_{0,j})/C_{0,i}| \times 100$, where indices $j$ and $i$ correspond to the coarser and finer time-steps, respectively. As summarised in table~\ref{tab:time_sensitivity}, the predicted maximum mean concentration remains invariant at $C_{0,\max} = 1.4675$ up to four decimal places across the three smallest temporal increments ($\Delta T = 10^{-2}$, $10^{-3}$, and $10^{-4}$). Discrepancies between these temporal levels only manifest beyond the sixth decimal place, resulting in negligible relative errors of $e_{\mathrm{rel}}^{32} < 10^{-4}\%$ and $e_{\mathrm{rel}}^{21} < 10^{-4}\%$. Both the computed values and the convergence behaviour shown in figure~\ref{fig:mesh_convergence} confirm that choosing $\Delta T = 10^{-3}$ is sufficient to guarantee temporal independence of the numerical solution.

\begin{table}
\centering
\begin{tabular}{ll}
\hline
Parameter & Value \\
\hline
$\Delta T_{1}, \Delta T_{2}, \Delta T_{3}$   & $10^{-4}, 10^{-3}, 10^{-2}$  \\
$C_{0,1}$ for $\Delta T_{1}$               & $1.4675$ \\
$C_{0,2}$ for $\Delta T_{2}$               & $1.4675$ \\
$C_{0,3}$ for $\Delta T_{3}$               & $1.4675$ \\
$e_{\mathrm{rel}}^{32}$                     & $< 10^{-4}\%$ \\
$e_{\mathrm{rel}}^{21}$                     & $< 10^{-4}\%$ \\
\hline
\end{tabular}
\caption{Relative error calculations for the peak concentration, $C_{0,\max}$, as a function of time-step on the fine mesh ($m = 1,000$).}
\label{tab:time_sensitivity}
\end{table}

Once temporal independence was established at $\Delta T = 10^{-3}$, the spatial discretisation uncertainty of the numerical scheme was assessed using the Grid Convergence Index (GCI), following the standardised guidelines of \citet{celik2008procedure} and \citet{roache1998verification}. The monitored quantity was selected as the maximum mean concentration, $C_{0,\max}$, at $T = 100$. Three mesh configurations were employed: coarse ($m_3 = 300$), medium ($m_2 = 500$), and fine ($m_1 = 1,000$), where $m_i$ denotes the total number of nodes along the axial coordinate $Z$.

The characteristic grid size for a uniform one-dimensional mesh is $h = \Delta Z = 1/(m_i - 1)$, where $m_i$ is the number of axial nodes. The refinement factors between the fine ($h_1$), medium ($h_2$), and coarse ($h_3$) grids are given by:
\begin{equation}
    r_{21} = \frac{h_2}{h_1} = 2.0000, \qquad r_{32} = \frac{h_3}{h_2} = 1.6722.
\end{equation}

Defining the solution differences as $\epsilon_{32} = \phi_3 - \phi_2$ and $\epsilon_{21} = \phi_2 - \phi_1$, the apparent order of convergence, $p$, is determined as follows:
\begin{equation}
    p = \frac{\ln | \epsilon_{32} / \epsilon_{21} |}{\ln(r_{21})}.
\end{equation}

The continuum solutions in the limit of zero grid spacing ($h \to 0$) are estimated using Richardson extrapolation:
\begin{equation}
    \phi_{\mathrm{ext}}^{21} = \frac{r_{21}^p \phi_1 - \phi_2}{r_{21}^p - 1}, \qquad \phi_{\mathrm{ext}}^{32} = \frac{r_{32}^p \phi_2 - \phi_3}{r_{32}^p - 1}.
\end{equation}

The approximate relative errors, $e_a$, and extrapolated relative errors, $e_{\mathrm{ext}}$, are computed as follows:
\begin{equation}
    e_a^{ji} = \left| \frac{\phi_i - \phi_j}{\phi_i} \right|, \qquad e_{\mathrm{ext}}^{ji} = \left| \frac{\phi_{\mathrm{ext}}^{ji} - \phi_i}{\phi_{\mathrm{ext}}^{ji}} \right|,
\end{equation}
where $(j, i) \in \{(3, 2), (2, 1)\}$. Finally, the Grid Convergence Index (GCI), with a safety factor of $F_s = 1.25$, is defined as follows:
\begin{equation}
    GCI_{\mathrm{fine}}^{21} = \frac{1.25 \, e_a^{21}}{r_{21}^p - 1}, \qquad GCI_{\mathrm{medium}}^{32} = \frac{1.25 \, e_a^{32}}{r_{32}^p - 1}.
\end{equation}

The apparent order of convergence, $p = 2.1495$, is consistent with the formal second-order accuracy expected from the central finite-difference discretisation employed in the Crank–Nicolson scheme. Furthermore, the Richardson continuum estimates at zero grid spacing yield $\phi_{\mathrm{ext}}^{32} = 1.4669$ and $\phi_{\mathrm{ext}}^{21} = 1.4674$. The latter is in close agreement with the fine-grid prediction ($\phi_1 = 1.4675$), corresponding to an extrapolated relative error of only $e_{\mathrm{ext}}^{21} = 0.0094\%$. In addition, the approximate relative error decreases from $e_a^{32} = 0.1436\%$ to $e_a^{21} = 0.0324\%$ with progressive spatial refinement.

The resulting medium-grid convergence index, $GCI_{\mathrm{medium}}^{32} = 0.0889\%$, indicates that the computations reside within the asymptotic convergence regime. Progression to the fine mesh further reduces the uncertainty index to $GCI_{\mathrm{fine}}^{21} = 0.0118\%$, confirming that the solution with $m_1 = 1,000$ deviates by only approximately $0.01\%$ from the extrapolated continuum state. This minimal uncertainty is consistent with the flat profile observed across all temporal levels in figure~\ref{fig:mesh_convergence} for grid densities exceeding 400 nodes. Accordingly, a spatial resolution of $m = 1,000$ nodes, combined with a time-step of $\Delta T = 10^{-3}$, is sufficient to ensure that the reported numerical results are independent of both spatial and temporal discretisations.

\begin{table}
\centering
\begin{tabular}{ll}
\hline
Parameter & Value \\
\hline
$N_1, N_2, N_3$              & $999, 499, 299$ \\
$r_{21}$                     & $2.0000$ \\
$r_{32}$                     & $1.6722$ \\
$\phi_1$ ($m_1 = 1000$)      & $1.4675$ \\
$\phi_2$ ($m_2 = 500$)       & $1.4670$ \\
$\phi_3$ ($m_3 = 300$)       & $1.4649$ \\
$p$                          & $2.1495$ \\
$\phi_{\mathrm{ext}}^{32}$     & $1.4669$ \\
$e_{a}^{32}$                 & $0.1436\%$ \\
$e_{\mathrm{ext}}^{32}$        & $0.0712\%$ \\
$GCI_{\mathrm{medium}}^{32}$   & $0.0889\%$ \\
$\phi_{\mathrm{ext}}^{21}$     & $1.4674$ \\
$e_{a}^{21}$                 & $0.0324\%$ \\
$e_{\mathrm{ext}}^{21}$        & $0.0094\%$ \\
$GCI_{\mathrm{fine}}^{21}$     & $0.0118\%$ \\
\hline
\end{tabular}
\caption{Computations of the Grid Convergence Index (GCI) for the maximum mean concentration $C_{0,\max}$ at $T = 100$.}
\label{tab:grid_independence_C0}
\end{table}

\subsection{Numerical checks and moment diagnostics}
\label{sec:axisymmetric_validation}
\label{sec:extended_validation}
\label{sec:long_time_refinement}

  Spatial refinement of the reduced model for
$(n, \Lambda) = (1, 0.05)$ and $(0.4, 0.2025)$,
with $\sigma_0 = 0.06$, altered the normalised pulse
width by less than $0.0061\%$.

For the latter case, the corresponding change
in the axisymmetric model was less than $0.019\%$;
time-step effects were even smaller.
The reversal near $T = 72.8$ persisted under refinement.

These convergence checks support the accuracy of the reference concentration histories but do not independently resolve the minor contribution of wall-induced redistribution to the pulse width.

Critical pulse widths are determined by identifying the roots of the integral variance growth rate, evaluated for a prescribed Gaussian concentration profile.

Pulse moments are evaluated using cross-sectional
and axial quadrature, and minimum widths and reversal
times are extracted from concentration histories
sampled every $0.1$ up to $T = 100$.

The parameter boundaries delineating net contraction
from recorded reversal are locally refined; colour
interpolation is restricted to cases exhibiting a recorded
reversal.

A minimum occurring at the final observation time
is not classified as a reversal, and minima occurring during
$T < 1$ are excluded from the reported reversal times.

The solute-retention and boundary-distance criteria described in
section~\ref{sec:pulse_moments} are applied throughout the analysis.

\FloatBarrier
\section{Prescribed-profile rheological criterion}\label{app:profile_criterion}
\label{sec:rheology_compression_criterion}

The interior variance balance can be reformulated to elucidate how power-law rheology modulates the competition between
advective compression and dispersive spreading. Specifically, we consider a prescribed axial concentration profile,
which is held fixed while $Pe$ varies at constant $n$, $\Lambda$, and $\epsilon$. We introduce
\begin{equation}
D_{\mathrm{eff}}=1+Pe^2d_n,\qquad
\mathcal K_w=Pe^2\kappa_n,\qquad
d_n=f(n)\langle U\rangle^2,\qquad
\kappa_n=\frac{n\Lambda}{2(5n+1)}P\langle U\rangle,
\label{eq:rheology_factorization}
\end{equation}
where $f(n)$ is defined in section~\ref{sec:limiting_cases}. The interior half-variance growth rate can then be expressed as
\begin{equation}
\frac{\mathcal F_{\mathrm{int}}}{\epsilon^2}
=1-\beta Pe+\alpha Pe^2,
\qquad
\frac12\frac{\mathrm d\sigma_Z^2}{\mathrm dT}
=\mathcal F_{\mathrm{int}}+\mathcal B_{\partial},
\label{eq:rheology_quadratic_budget}
\end{equation}
with $\mathcal F_{\mathrm{int}}=\mathcal B_U+\mathcal B_w+\mathcal B_D$ and
\begin{equation}
\begin{aligned}
\beta&=-\epsilon^{-1}\bigl\langle(Z-\mu)\langle U\rangle\bigr\rangle_C,\\
\alpha&=\alpha_D+\alpha_w,\\
\alpha_D&=\left\langle d_n+(Z-\mu)\frac{\mathrm{d}d_n}{\mathrm{d}Z}\right\rangle_C,\qquad
\alpha_w=-\langle(Z-\mu)\kappa_n\rangle_C.
\end{aligned}
\label{eq:rheology_competition_coefficients}
\end{equation}
These expressions retain the contribution from the axial gradient of the effective dispersion coefficient without invoking the narrow-pulse approximation. For a prescribed concentration profile and fixed hydrodynamic parameters, $\alpha$ and $\beta$ are independent of $Pe$. In the positive outlet throughflow regime with $\Lambda > 0$, the downstream decrease in mean axial velocity ensures $\beta > 0$ for any non-degenerate pulse. Similarly, the monotonicity established in~\eqref{eq:kw_monotone} implies $\alpha_w \geq 0$. No general sign restriction is imposed on $\alpha_D$.

For $\alpha > 0$ and negligible axial boundary contributions, instantaneous pulse contraction arises over a finite interval of $Pe$ if and only if
\begin{equation}
\Theta\equiv\frac{\beta^2}{4\alpha}>1.
\label{eq:rheology_compression_number}
\end{equation}
Within the retained model, the corresponding interval is
\begin{equation}
Pe_-<Pe<Pe_+,\qquad
Pe_{\pm}=\frac{\beta\pm\sqrt{\beta^2-4\alpha}}{2\alpha}.
\label{eq:rheology_pe_interval}
\end{equation}
  At $\Theta = 1$, the interior variance growth rate is zero at a single value of $Pe$ but remains non-negative elsewhere. These roots represent formal predictions for the family of prescribed profiles and do not imply that the homogenised model remains accurate for arbitrarily large $Pe$. Only those regions lying within the model's regime of validity can be interpreted quantitatively. During pulse evolution, the profile, centroid, and these weighted coefficients change; thus, equation~\eqref{eq:rheology_compression_number} does not provide a permanent classification for a given pulse.

For a Gaussian distribution with $Z_0 = 0.2$, $\sigma_0 = 0.04$, $\epsilon = 0.0025$, and $\Lambda = 0.223$, decreasing $n$ from 1 to 0.4 alters $\beta$ only marginally, while it substantially reduces $\alpha$ and, consequently, increases $\Theta$. For this prescribed profile, the rheological dependence of $\Theta$ is therefore governed primarily by the reduction in the coefficient of the quadratic $Pe^2$ term, rather than by an increase in the magnitude of the compressive covariance.

The wall correction increases $\alpha$ by $\alpha_w$ for a fixed flow and prescribed profile. For $\alpha_D > 0$, it therefore reduces $\Theta$ and narrows any compression interval that would exist in the absence of this correction. This analytical observation distinguishes the opposing effect of wall-induced concentration redistribution from the compressive contribution of mean-flow deceleration.

\end{appen}

\bibliographystyle{jfm}
\bibliography{jfm}



\end{document}